\documentclass[11pt, preprint]{aastex}
\usepackage{hyperref}
\usepackage{float}
\usepackage{subfigure}
\usepackage{placeins}
\usepackage{morefloats}

\newcommand{\noprint}[1]{}
\newcommand{\figsetstart}{{\bf Fig. Set} }
\newcommand{\figsetend}{}
\newcommand{\figsetgrpstart}{}
\newcommand{\figsetgrpend}{}
\newcommand{\figsetnum}[1]{{\bf #1.}}
\newcommand{\figsettitle}[1]{ {\bf #1} }
\newcommand{\figsetgrpnum}[1]{\noprint{#1}}
\newcommand{\figsetgrptitle}[1]{\noprint{#1}}
\newcommand{\figsetplot}[1]{\noprint{#1}}
\newcommand{\figsetgrpnote}[1]{\noprint{#1}}

\begin{document}
\title{Young Stellar Populations in MYStIX Star Forming Regions: Candidate Protostars}
\author{Gregory Romine\altaffilmark{1}, Eric D. Feigelson\altaffilmark{*1,2,3}, Konstantin V. Getman\altaffilmark{1}, Michael A. Kuhn\altaffilmark{3,4}, Matthew S. Povich\altaffilmark{5}}

\altaffiltext{*} {edf@astro.psu.edu} 
\altaffiltext{1}{Department of Astronomy \& Astrophysics, Pennsylvania State University, 525 Davey Lab, University Park, PA 16802}
\altaffiltext{2}{Center for Exoplanets and Habitable Worlds}
\altaffiltext{3}{Millennium Institute of Astrophysics}
\altaffiltext{4}{Instituto de Fisica y Astronom\'ia, Universidad de Valpara\'iso, Gran Breta\~na 1111, Playa Ancha, Valpara\'iso, Chile}
\altaffiltext{5}{Department of Physics and Astronomy, California State Polytechnic University, 3801 West Temple Ave, Pomona, CA 91768}

\slugcomment{Accepted for publication in AAS Journals September 2016}

\keywords{infrared: stars $-$ ISM: individual objects(W 3, Flame Nebula, Rosette Nebula, NGC 2264, RCW 38, RCW 36, NGC 6334, NGC 7357, Trifid Nebula, Lagoon Nebula, M 17, Eagle Nebula, W 40, DR 21) $-$ protoplanetary disks $-$  stars: formation $-$ stars: protostars $-$ X-rays: stars}

\begin{abstract}
The Massive Young Star Forming Complex in Infrared and X-ray (MYStIX) project provides a new census on stellar members of massive star forming regions within 4~kpc.   Here the MYStIX Infrared Excess catalog (MIRES) and $Chandra$-based X-ray photometric catalogs are mined to obtain high-quality samples of Class~I protostars using criteria designed to reduce extragalactic and Galactic field star contamination.  A total of 1,109 MYStIX Candidate Protostars (MCPs) are found in 14 star forming regions.  Most are selected from protoplanetary disk infrared excess emission, but 20\% are found from their ultrahard X-ray spectra from heavily absorbed magnetospheric flare emission.  Two-thirds of the MCP sample is newly reported here.     The resulting samples are strongly spatially associated with molecular cores and filaments on \textit{Herschel} far-infrared maps.  This spatial agreement and other evidence  indicate that the MCP sample has high reliability with relatively few 'false positives' from contaminating populations.  But the limited sensitivity and sparse overlap among the infrared and X-ray subsamples indicate that the sample is  very incomplete with many 'false negatives'.   Maps, tables, and source descriptions are provided to guide further study of star formation in these regions.  In particular, the nature of ultrahard X-ray protostellar candidates without known infrared counterparts needs to be elucidated. 
\end{abstract}

\section{Introduction: Photometric identification of protostars}

In the densest cores of molecular clouds, gravitational collapse overcomes thermal pressure, turbulence, and rotational and magnetic energies to form protostars.  The protostellar phases of pre-main sequence evolution, lasting an estimated 0.5~Myr, are characterized by infall of material from the circumstellar envelope, formation of a protoplanetary disk, and bipolar outflows collimated by magnetic fields \citep{Dunham14}.  Protostars can be identified in three primary ways:  \begin{enumerate}

\item A strong photometric infrared excess due to thermal emission from the envelope and disk dust grains discriminates protostars from less interesting foreground and background stars. The infrared spectral energy distribution (SED) provides an evolutionary classification classification \citep{Adams87, Andre00}.  Class~0 protostars appear in the far-infrared, too obscured to be detected in the near- and mid-infrared bands.  Class~I protostars have a positive spectral index over the $2-25 \mu m$ band.   A flat spectral index is sometimes classified as a Class I/II transition,  Class~II stars have a mildly negative spectral index, and stars without disks exhibiting the Rayleigh-Jeans slope are classified as Class III \citep{White07}.  The SED may depend on the chance orientation of the disk; for example, a Class II T Tauri star with disk seen edge-on can mimic Class I protostellar SED \citep{Whitney03}.   Furthermore, protostellar infrared colors can be similar to contaminating cosmic populations: dust-enshrouded AGB post-main sequence stars, starburst galaxies, Type 2 obscured active galactic nuclei, and shocked interstellar gas emission.  Cuts in color-color and color magnitude diagrams have been developed help discriminate protostars from these contaminants \citep{Gutermuth09}. 

\item Protostars are distinguished by strong outflows that are directly traced by molecular maps showing redshifted and blueshifted lobes.  When molecular maps are not available, infrared line emission from outflow shocks can sometimes be detected photometrically.  The {\it Spitzer Space Telescope} 4.5 $\mu m$ band contains excited H$_2$ lines and CO band heads that can produce a distinctive photometric [4.5] excess \citep{Cyganowski08}. 

\item Throughout the Class I-II-III stages of pre-main sequence evolution, low mass stars show very high levels of surface magnetic activity that are most easily detected by X-ray flaring \citep{Feigelson99}. X-ray flares from protostars and T Tauri stars are similar to, but much stronger and more frequent than, magnetic reconnection events flares above the surface of the contemporary Sun.   The high extinction of deeply embedded protostars absorbs softer energy photons, leaving only harder X-ray photons; spectral hardness is thus a signature of Class~I stars \citep{Imanishi01, Ozawa05}.   The median energy $ME$ (in keV) of detected photons from a young star is a reliable measure of soft X-ray absorption \citep{Getman10}.   Class~0 stars have never been detected in X-rays, but the limits are consistent with Class~I levels subject to  heavy absorption \citep{Giardino07}.  

\end{enumerate}

Infrared excess protostellar samples are readily obtained from imaging surveys obtained with the \textit{Spitzer Space Telescope} and \textit{Herschel Space Observatory} in nearby smaller star forming clouds \citep{Evans03, Andre10} and the nearest Orion giant molecular cloud \citep{Megeath12, Stutz13}.  Less complete samples of Class~I protostars are obtained from X-ray surveys with the {\it Chandra X-ray Observatory} and {\it X-ray Multimirror Mission} in nearby clouds such as $\rho$ Ophiuchi \citep{Casanova95, Imanishi01}, Corona Australis \citep{Koyama96, Hamaguchi05}, and Serpens \citep{Preibisch04, Winston07}, as well as the Orion giant molecular cloud \citep{Grosso05, Prisinzano08, Pillitteri13}.  

However, it is difficult to obtain unbiased flux-limited protostellar samples in more distant and massive Galactic star forming clouds for three reasons.  First, the protostellar infrared sources are fainter and more crowded, difficult to resolve from neighboring  protostars or the more common Class~II stars in embedded clusters.  Second, more distant star forming regions are concentrated in the Galactic Plane where contamination by dusty post-main sequence stars can be severe.  The surface density of field stars in $Spitzer$ images at low Galactic latitudes and longitudes can reach the confusion limit fainter than $\sim 12$~mag in short exposures due to red giant field stars \citep{Ramirez08}.  Third, in the vicinity of H\textsc{ii} regions ionized by recently formed OB stars, unresolved protostellar sources can be difficult to detect within the spatially complex diffuse emission from heated dust, especially in infrared bands with strong polycyclic aromatic hydrocarbon emission.  In some cases, reasonable protostellar samples can be obtained despite diffuse emission, as in the Rosette Molecular Cloud \citep{Hennemann10}, but in other cases only luminous high-mass protostars can be identified, as in the NGC~6334 complex \citep{Brogan09}.   
 
A new opportunity for the discovery of protostars in giant molecular clouds emerges from the Massive Young Star-forming complex in Infrared and X-ray (MYStIX) project.  MYStIX combines infrared and X-ray photometric surveys of 20 OB-dominated young star forming complexes at distances from 0.5 to 4 kpc.  MYStIX uses spectral-imaging data from NASA's  \textit{Chandra X-ray Observatory}, $1-2$~$\mu$m images from the UK InfraRed Telescope (UKIRT) wide-field camera, and $3-8$~$\mu$m images from NASA's \textit{Spitzer Space Telescope} \citep{Feigelson13}. This survey produced a catalog of 31,784 MYStIX Probable Complex Members (MPCM) in the 20 regions \citep{Broos13}. Most MPCM stars are Class~II and Class III pre-main sequence stars, but a small fraction should be Class~I protostars.  The much larger catalogs of unmatched MYStIX Infrared-Excess Source \citep[MIRES;][]{Povich13}  catalog and X-ray sources \citep{Kuhn13a, Townsley14} produced as intermediate products during the construction of the MPCM catalog may contain additional protostars, although these single-waveband catalogs suffer strong contamination by Galactic field stars and/or extragalactic sources.  

The present study seeks protostars from these MYStIX survey catalogs.  We combine objects with protostellar infrared SEDs and 4.5~$\mu$m excesses with X-ray sources exhibiting ultra-hard spectra denoting very heavy obscuration.  These criteria filter away nearly all of the older Class II-III stars and contaminant populations, but give very incomplete samples.  The result is a list of 1,109 protostellar candidates in fourteen star forming regions (in order of increasing right ascension): W~3, Flame Nebula, Rosette Nebula, NGC~2264, RCW~38, RCW~36, NGC~6334, NGC~6357, Trifid Nebula, Lagoon Nebula, M~17, Eagle Nebula, W~40, and DR~21.   A fifteenth MYStIX region with the 5 Myr cluster NGC~2362, was chosen as a control field as no protostars are expected in this region without molecular material.  The reliability of the catalog is strengthened because a large majority (86\%) are found to be associated with dense cores seen in \textit{Herschel} 500~$\mu$m maps that trace cold dust emission.  However, the candidate list requires more detailed study for confirmation, and cannot be viewed as an unbiased view of star formation in the clouds.  

Section 2 details the methods of selection for the candidate protostars, sections 3-4 present results for individual star forming regions, and section 5 discusses the overall findings.

\section{Protostar selection procedures} \label{selection.sec}

\subsection{The MYStIX catalogs} \label{mystix.sec}

The MYSTtIX effort\footnote{\citet{Feigelson13} gives an overview of the MYStIX procedure, electronic tables of intermediate MYStIX catalogs are published with papers detailed various stages of the analysis, and the final MPCM catalog accompanies \citet{Broos13}.}  starts with analysis of archival $Chandra$ observations of OB-dominated young stellar clusters and their molecular clouds with advanced software that detects sources as faint as $3-5$ counts with high reliability \citep{Kuhn13a, Townsley14}.  Each $Chandra$ observation subtends $17\arcmin \times 17\arcmin$\/ and most fields have two or more overlapping pointings.   A total of 34,530 faint X-ray sources is located in 20 MYStIX regions.  Wide-field $JHK$ surveys of these regions with the UKIRT WFCAM instrument \citep[some observations are from the UKIDSS Galactic Plane survey,][]{Lucas08}), as well as $3.5-8.0$~$\mu$m surveys obtained with $Spitzer$'s IRAC instrument, are reanalyzed with emphasis on detection and photometry in crowded and nebulous environments \citep{King13, Kuhn13b}.   These catalogs have 1,056,771 near-infrared and 333,406 mid-infrared sources in 18 regions. The Orion Nebula and Carina Nebula were omitted from these catalogs because they were analyzed by our group's pre-MYStIX studies.  Clearly the vast majority of infrared sources are not associated with the star forming regions.  

Matching of X-ray sources to infrared stars is performed in a probabilistic magnitude-dependent fashion to reduce spurious matches to members of the populations of uninteresting older field  stars in the infrared images or to extragalactic active galactic nuclei in the X-ray images \citep{Naylor13}.  Matching probabilities are affected by the strong gradient in point spread function of the $Chandra$ mirrors that gives sub-arcsecond accuracy close to the pointing direction but several arcseconds accuracy near the detector edge.  

Near- and mid-infrared spectral energy distributions of infrared sources are then calculated, and decision rules are applied to discriminate disk-bearing young stellar objects from dusty AGB stars and starburst galaxies \citep{Povich13}.  These rules are considerably stricter than those usually applied in studies of nearby star forming regions away from the Galactic Plane due to the high surface density of contaminating populations.  Povich et al.\ produce a catalog called MYStIX InfraRed Excess Sources (MIRES) based on these criteria.  Subsets showing ascending SEDs and/or [4.5]$\mu$m excess emission associated with protostars are noted in this catalog.

A naive Bayes classifier then combines several criteria in a probabilistic fashion to define the list of MYStIX Probable Complex Members \citep[MPCM;][]{Broos13}.  The criteria include: MIRES sources with SEDs resembling Class~I or Class~II young disk-bearing stars; MIRES sources with [4.5] excesses; X-ray sources with counterparts having $J$ magnitudes based on a training set of strongly clustered sources in that region; X-ray sources having median energies characteristics of absorption in that region; X-ray sources exhibiting rapid flares; and spectroscopically confirmed OB stars.  Broos et al.\ emerge with 31,784 MYStIX Probable Complex Members (MPCMs).  

Science analysis of the MPCM catalog was pursued in several later studies.  First, the MPCM stars in 17 of the star forming regions were segmented into 142 (sub)clusters representing spatial concentration of clusters using a mixture model of clusters plus a distributed population fitted by a maximum likelihood procedure \citep{Kuhn14}.  Second, total stellar populations of the subclusters were estimated by normalizing the observed X-ray luminosity functions, based on $Chandra$ observations with different sensitivities, to the X-ray luminosity function of the well-studied Orion Nebula Cluster \citep{Kuhn15a}.  Third, age estimates of the stars in most of these subclusters were obtained using a new technique based on X-ray and infrared photometry \citep{Getman14}. These $Age_{JX}$ estimates show clear age gradients across the star forming regions from youngest embedded clusters to intermediate-age revealed clusters to oldest distributed pre-main sequence populations.  

We treat 14 of the 20 MYStIX target regions \citep{Feigelson13} in this study.  Two of the regions have already been carefully examined for protostars using multiwavelenth X-ray/infrared techniques:  the Carina Nebula \citep{Smith10, Povich11} and the Orion Nebula \citep{Grosso05, Prisinzano08}.  Note that these fields were not examined for ultrahard X-ray sources (\S\ref{ultrahardX.sec}).  We further require that $Herschel$ satellite far-infrared dust maps be available to associate the candidate protostars with dense molecular cores and filaments (\S\ref{Herschel_cores.sec}).  This criterion leads to the omission of W~4, NGC~3576 and NGC~1893, and NGC~2362.  However, this last region is examined as a control field (\S\ref{NGC2362.sec}).

\subsection{Ultrahard X-ray sources} \label{ultrahardX.sec}

\subsubsection{Selection criteria} \label{Xsel.sec}

It has long been recognized that interstellar gas is an effective absorber of soft X-rays \citep{Brown70, Wilms00}.  For an assumed normal gas-to-dust ratio and solar abundances of the elements, the line-of-sight column density $N_H$ derived from X-ray absorption is approximately linearly related to visible absorption $A_V$ in magnitudes according to $N_H \simeq 2 \times 10^{21}~A_V$~cm$^{-2}$, although the proportionality constant is uncertain over the range $1.8-2.8 \times 10^{21}$ cm$^{-2}$ \citep{Vuong03, Coley15}.   Line-of-sight absorption to young stellar objects can thus be estimated from X-ray absorption.

\citet{Getman10} has shown that, for faint X-ray sources too faint for parametric spectral modeling, the simple measure of the median energy $ME$ (in keV) from a CCD spectrum of a young star is a reasonably good measure of soft X-ray absorption and $N_H$.  An intrinsic spectrum for a chosen X-ray luminosity and the characteristics of a particular X-ray telescope and instrument must be specified to calibrate observed $ME$ values to $N_H$ values.  For the $Chandra$ ACIS instrument and a pre-main sequence star with $L_x \sim 10^{30}$ erg~s$^{-1}$,  the correspondence is roughly as follows:  for $ME =  2.0$, 3.0, 4.0, 5.0, and 6.0 keV, the inferred line-of-sight column density is $\log N_H \simeq 22.1$, 22.5, 23.0, 23.5, and 23.9 cm$^{-2}$ (see Getman et al., Figure 4).  Assuming a gas-to-dust calibration of $N_H = 2 \times 10^{21}~A_V$~cm$^{-2}$, these median energies correspond to $A_V \simeq  6$, 16, 50, 150, and 400 mag.  The nonparametric median energy parameter can be estimated with reasonable accuracy for X-ray sources that are too faint for parametric spectral modeling.  

The MPCM catalog suffers an important limitation with respect to protostars; it is called the `likelihood tail problem' by \citet[][\S6.3]{Broos13}.  Their `naive Bayes' classifier of MYStIX X-ray sources used to construct the MPCM sample assumes that the probability density functions for the properties $-$ such as X-ray median energy of young stellar objects and of all classes of contaminant populations $-$ are known perfectly.   But due to finite sample sizes, the tails of these distributions are quite uncertain, and the classifier behavior will be unstable.  Broos et al.\ state: `[A]t the hard end of the median X-ray energy distributions (Figure 2), e.g., above $\sim$4.5 keV; sources this hard are very rare in all classes. Protostars in the [massive star forming regions] are expected to lie here, and we should expect that the classifier will have difficulty distinguishing them from contaminants.'   This problem motivates our use of ultra-hard X-ray sources in the original X-ray source catalogs, irrespective of whether infrared counterparts are present or whether the naive Bayes classifier considered them to be probable young stars (i.e., MPCM members).  

After some experimentation, we developed here a dual criterion for uncovering candidate protostars from \textit{Chandra} X-ray catalogs in the MYStIX fields \citep{Kuhn13a, Townsley14}.  First, we set a conservative median energy cut $ME > 4.5$~keV that corresponds to a hydrogen column density $N_H \approx 2 \times 10^{23}$ (assuming solar abundances in the intervening gas) or $A_v > 100$ mag (assuming standard interstellar gas-to-dust ratio in the intervening material) using the calibration og \citet{Getman10}. The threshold $ME > 4.5$~keV was chosen partly from past studies of star forming regions where previously known Class I protostars were found to have $ME \geq 4$~keV \citep[e.g.,][]{Getman07, Schneider09, Principe14}, and partly from examination of the spatial distribution of sources in MYStIX fields under more and less strict criterion. Lower criterion like $ME>3.5$ or $ME>4.0$ increases the number of X-ray sources in the fields that are randomly located in the field rather than concentrated around dense molecular cloud cores (\S\ref{Herschel_cores.sec}) and increases the number of sources seen in the older field NGC~2362 where no Class 0/I protostars are expected. 

Candidate protostars were further culled by requiring that at least 10 net counts are present in the X-ray source.  Net counts are the number of events extracted from the local point spread function minus the number expected from local background.  This criterion is considerable higher than the MYStIX detection limit, typically $3-5$ net counts on-axis.  As with the median energy criterion, we found that reducing the count threshold increase the number of X-ray sources in the fields that were randomly located in the field and the number of sources in the NGC~2362 field.  For example, a criterion of $>5$ net counts with $ME > 4.5$~keV gives four candidate protostars in NGC~2362, that is reduced to one candidate when $>10$ net counts is required.   This count requirement also removes sources where the statistical uncertainty of the median energy is high, $\Delta ME > 0.5$~keV \citep[see Figure 9 of][]{Getman10}. 

\subsubsection{Extragalactic contamination} \label{AGN.contam.sec}

Our sample of ultra-hard X-ray sources should suffer contamination from heavily absorbed active galactic nuclei (AGN). In these objects, known as Type 2 AGN, the X-ray emission from the accretion process near the supermassive black hole is obscured by a local dusty torus.  Type 2 AGN are roughly as common as less absorbed Type 1 AGN, and together they dominate the faint X-ray source population in deep X-ray images of the extragalactic sky.  

We can roughly estimate the level of this contaminating population as follows.  A typical $Chandra$ exposure in  MYStIX fields have duration around 100~ks \citep[Table 2,][]{Feigelson13}, although some fields are several times deeper or more shallow.  A typical AGN has power law X-ray spectrum with photon index $\Gamma = 1.8$.  For our limiting count rate of 10 counts in 100~ks and an assumed absorption around $\log N_H \simeq 23$~cm$^{-2}$ for a Type 2 AGN, the limiting AGN flux in the $2-10$~keV band is $\simeq 4 \times 10^{-15}$ erg s$^{-1}$ cm$^{-2}$ as calculated with NASA's {\it Portable, Interactive Multi-Mission Simulator}\footnote{ \url{http://heasarc.gsfc.nasa.gov/docs/software/tools/pimms.html} }.  

From a collection of AGN flux-limited samples observed with the $Chandra$ and $X-ray~ Multi-Mirror$ telescopes, \citet{LaFranca05} measure the sky source density of absorbed AGN as a function of limiting flux and absorption.  Combining their source count distribution (Fig.\ 15) and $\log N_H$ distribution (Fig.\ 10) for our flux limit, we estimate that $\sim 30$ deg$^{-2}$ Type 2 AGN should be present in our MYStIX X-ray images with $ME > 4.5$~keV and $> 10$ net counts.  As different MYStIX regions have different patterns of overlapping $Chandra$ fields with different exposure times, predicted Type 2 AGN contamination ranges roughly from $2-20$ objects per MYStIX region.  

The distinctive difference between faint ultra-hard protostellar and AGN populations should be their spatial distributions and infrared counterparts. Type 2 AGN should be randomly distributed across $Chandra$ fields, while protostars should be concentrated in star-forming cloud cores.  Type 2 AGN will always have extremely faint infrared counterparts while protostars may have bright infrared counterparts.  

\subsection{Infrared selected sources} \label{irexcess.sec}

The MIRES catalog constructed from the MYStIX near- and mid-infrared catalogs by \citet{Povich13} provides a sample of candidate protostars selected by the shape of the infrared SED. The MIRES catalog is based on UKIRT $JHK$ and $Spitzer$ IRAC [3.6]$-$[8.0] images that typically extend beyond the $Chandra$ fields used to define MYStIX regions. Sources are classified as designated 'envelope-dominated Stage 0/I' (or 'S0/I') sources when the following conditions are satisfied: (a) at least 4 of the 7 photometric bands ($JHK$ from UKIRT and 3.6, 4.5, 5.8 and 8.0~$\mu$m bands from $Spitzer$) are available; and (b) SEDs show  infrared excesses over stellar photospheres that are successfully fit to star-plus-disk-plus-envelope models of \citet{Robitaille07}. Unreliable cases are excluded according to three rules: the infrared excess is confined to a single $Spitzer$ IRAC band (and thus possibly inaccurate photometry due to crowding or nebular contamination); anomalous blue [5.8]$-$[8.0] colors are present; or the [3.6]$-$[4.5] color is consistent with interstellar reddening. The MIRES catalog has 2730 infrared sources classified as S0/1 of which 1646 are located within $Chandra$ fields of view (see Tables 5-6 in Povich et al.). 

The spatial distribution of S0/1 sources show that many are randomly distributed across the MYStIX fields, rather than being concentrated near cloud cores, indicating strong contamination by dusty AGB stars, H\textsc{ii} region nebular knots and/or starburst galaxies. We therefore applied two additional constraints applied to the S 0/I sources:

\begin{description}

\item[Color selected sources]  For MYStIX sources that are classified as `Stage 0/I' from SED fitting to protostellar models by \citet{Povich13}, we applied the $Spitzer$ IRAC color-color classification procedure described by \citet{Gutermuth09} to discriminate `Class I' protostars from other objects.  The procedure defines a polygonal region in the [3.6]-[4.5] $vs.$ [4.5]-[5.8] diagram; see `Phase 1' in their Appendix. These selection criteria first remove most starburst galaxies, active galactic nuclei, unresolved shock emission, and interstellar PAH contamination. 

\item[4.5 $\mu$~m excess sources]  Based on the study by \citet{Cyganowski08},  we apply a constraint for excess 4.5 $\mu$m emission in addition to a blackbody disk emission.  This is caused by shocked molecular emission in protostellar outflows and should be a high quality selector for a fraction of protostars.  Section \S3.3 and  Figure 3 of \citet{Povich13} describe this criterion and the subset of MIRES sources that exhibit the excess emission. 

\end{description}

We find that 710 MIRES sources satisfy the infrared color constraint (designated `IR' in Tables~1 and 2), and 295 satisfy the [4.5] excess constraint (designated `4E' in Tables~1 and 2.  Any MIRES star that satisfy one or both of these constraints is included in our MYStIX candidate protostar list here. 

\subsection{Dense molecular cloud cores} \label{Herschel_cores.sec}

Most protostars are expected to reside in or close to their natal cold molecular gas that can be traced with thermal emission from cold dust is in the 70-500 $\mu m$ range \citep{Molinari13}.  We there compared the location of candidate protostars obtained from the infrared and X-ray criteria described above with  maps obtained with the \textit{Herschel Space Observatory} Spectral and Photometric Imaging Receiver (SPIRE)  at 500$\mu$m.  Note that location in or near a cloud core or filament is not a basis for selection, but rather is used for scientific interpretation.  Candidates in or near \textit{Herschel} clouds and filaments are more likely to be true protostars  than those found outside the cold dust emission that have a greater chance of being members of contaminant populations.

\section{Results} \label{results.sec}

\subsection{Presentation of MYStIX Candidate Protostars} \label{present.sec}

Table~ \ref{summary.tbl} summarizes the sample of protostellar candidates found within the MYStIX fields.  The central results are given in the first four columns: the MYStIX field listed in order of right ascension, total number of MYStIX Candidate Protostars (MCPs) found, subsample found from infrared excess (IRE), and subsample found from X-ray emission found using the selection rules of \S\ref{ultrahardX.sec}-\ref{irexcess.sec}.  Note that the counts in column 2 are typically less than the sum of the counts in columns $3-5$, as some candidates are found with more than one selection procedure.  Columns $5-7$ give secondary characteristics of the samples: number of candidates associated with dense cores and filaments; number of candidates previously listed in our MPCM catalog \citep{Broos13}; and number of candidates previously identified  by other researchers.  The association with cores and filaments seen on $Herschel$ maps is a subjective evaluation: candidates on the edges of bright far-infrared cores, for example, are included but those superposed on less concentrated dusty regions are not.  The count of previously known protostars is similarly subjective, based on the {\bf individual object notes in} Table~\ref{MCPnotes.tbl}.  Sources previously found to have infrared excess but without notation of possible protostellar classification are not included in this count.  

Table~\ref{MCP.tbl} provides a source-by-source listing of the protostellar candidates in each MYStIX field with the following columns: 
\begin{description}

\item[Column 1: MCP designation] We define here `MCP [region]-\#' where MCP means `MYStIX Candidate Protostar', [region] represents the MYStIX star forming region (column 1 of Table~\ref{summary.tbl} without spaces), and \# is a running sequence number (up to values in column 2 of Table~\ref{summary.tbl}).    

\item[Column 2: MYStIX catalog designation]  Entries in galactic coordinates are from the MIRES catalog of \citet{Povich13}, and entries in celestial coordinates are from the X-ray catalogs of \citet{Kuhn13a} and \citet{Townsley14}.  

\item[Columns 3-4: Location]  These locations in J2000 celestial coordinates are from the corresponding catalog in column 2.  Note that MPCM positions (column 5) may be slightly offset ($<1\arcsec$) as they represent our best estimate from multiwavelength catalogs \citep{Broos13}.  MPCM positions take precedence over those given here, when available.

\item[Column 5: MPCM flag] Binary indicator whether the source appears in the MYStIX Probable Complex Member catalog \citep{Broos13}.  MPCM sources are located within the $Chandra$ field of view (yellow polygons in Figure~1) and satisfy a probabilistic decision tree procedure described by Broos et al.

\item[Column 6: MCP selection flag] Origin(s) of the source's selection as a MYStIX Candidate Protostar: IR indicates the source is obtained from the infrared excess criterion (\S\ref{irexcess.sec}), 4E indicates it is obtained from the [4.5] excess criterion (\S\ref{irexcess.sec}), and X indicates it obtained from the ultrahard X-ray source criteria (\S\ref{ultrahardX.sec}). 

\item[Column 7: Herschel flag] Binary indicator whether the source lies on or near a dense dusty core or filament in the $Herschel$ SPIRE 500$\mu$m map.

\item[Column 8: Subcluster identification]  For most sources in the MPCM catalog, this column gives the subcluster identifier from the maximum likelihood mixture model described by \citet{Kuhn14}.  A subcluster identifier (A, B, C, ...) indicates the source is part of a statistically significant grouping of MPCM stars, while a U identifier indicates the source lies in an unclustered environment. 

\item[Columns 9-12: $Chandra$ X-ray properties]  Obtained from the $Chandra$ catalogs of \citet{Kuhn13a} and\citet{Townsley14}, these columns give the median energy in keV of the net counts, the net counts (extracted photons minus local background), the probability of source non-existence (based on a Poisson model of the local background), and the probability of a constant X-ray count rate during the $Chandra$ exposure(s) (based on a Kolmogorov-Smirnov test).  Small values of these probabilities (e.g., $P<0.001$) denote highly significant detections with variable X-ray emission.  Note that the probability of source existence does not scale in a simple fashion with net counts due to variations in the point spread function across the $Chandra$ field of view. 

\item[Columns 13-16: $Spitzer$ infrared photometry] The [3.6], [4.5], [5.8], and [8.0] band magnitudes obtained by \citet{Kuhn13b}.  

\item[Column 17: Notes]  {\bf Object notes given in Table \ref{MCPnotes.tbl}} based on published studies.  These are derived from sources in the SIMBAD database\footnote{\url{http://simbad.u-strasbg.fr}} within 2\arcsec\/ of the location given in columns 3-4 and their associated papers.  Given the complexity of star forming regions and the heterogeneity of past studies, a physical association between the MYStIX source and published sources is not guaranteed.  

\end{description}

Table~\ref{summary.tbl} reports 1,109 MYStIX Candidate Protostars (MCPs) in the 14 MYStIX regions under study here.  Two-thirds of the MCPs (710 or 64\%) are found with infrared excess photometric criteria, a quarter (295 or 27\%) are found with [4.5]~$\mu$m excess characteristic of shocked outflows, and a quarter (253 or 23\%) are found as ultrahard X-ray sources.  One hundred thirty four are found using both infrared photometric criteria, and five are found using both X-ray and infrared criteria. Forty-seven infrared-selected candidates are also $Chandra$ X-ray sources that did not satisfy our X-ray selection criteria due to inadequate absorption ($ME < 4.5$~keV) or too few photons ($Cts < 10$), and 25 X-ray selected candidates are also $Spitzer$ infrared sources that did not satisfy our infrared color selection criteria. 

Visualization of the results is presented in Figure~1 where the the MYStIX protostellar candidates are superposed on $Herschel$ 500$\mu$m maps.  In these gray-scale maps, bright areas denote dense cores and filaments with strong thermal far-infrared emission.  The symbols discriminate protostellar candidates selected using standard infrared color criteria (red boxes), [4.5]~$\mu$m excess criterion (red crosses), and X-ray criteria (green circles). The yellow polygons show the fields of view of the $Chandra$ exposures in the region; each $Chandra$ field subtends $17\arcmin \times 17\arcmin$.  Note that infrared excess sources found using the criteria in \S\ref{irexcess.sec} can lie outside the $Chandra$ field because the MIRES catalog of \citet{Povich13} typically covers a larger area around the $Chandra$ fields.  However, the MPCM catalog of \citet{Broos13} is restricted to the $Chandra$ coverage areas.

\section{Individual star forming regions}  \label{individual.sec}
We summarize here information in Table \ref{MCP.tbl} and Figure~1 for each star forming region.

\subsection{W~3}

The W3 W4-W5 star forming complex is relatively nearby with the fortuitous location at $(l,b)=(134,1)$ that gives a reduced field star contamination compared to most MYStIX star formation regions that are projected onto the inner Galactic quadrants.  However, bright diffuse nebulosity pervades the entire region in mid-infrared bands, hindering identification of protostars by photometric infrared excess.  The W3 region alone has had several episodes of star formation producing a diversity of stellar clusters.  The most remarkable is the embedded cluster W3 Main that exhibits a rich spherical cluster of older pre-main sequence stars around collection of compact, ultracompact and hypercompact H\textsc{ii} regions from OB stars of various ages \citep{Tieftrunk97, Feigelson08, Bik14}.  The stellar population of the W3 region is previously studied at infrared wavelengths by various groups \citep{Ruch07, Rivera-Ingraham11, Bik14, Roman15} and at X-ray wavelengths by \citet{Feigelson08}. 

Forty one stars satisfy our criteria for protostellar candidates, all but two associated with  $Herschel$ cloud emission.  Within W3~Main in the central $Chandra$ fields, a dense concentration of ultrahard X-ray stars is associated with the W3~IRS5 cluster, including candidate \#15\footnote{For brevity in this section, we use the informal designations like `candidate \#15' to represent the formal designations like `MCP W3 15'.} associated with the $\sim 20$~M$_\odot$ O8 star IRS-MIR1 that ionizes the hypercompact H\textsc{ii} region W3~M.  Its X-ray emission is extremely heavily absorbed ($ME \simeq 5.6$~keV, $\log N_H \simeq 23.5-24.0$~cm$^{-2}$).  Other MCPs  are scattered around the W3~Main cluster; eight of these were found earlier by \citet{Rivera-Ingraham11} who report a total of 39 Class~0/I young stellar objects in W3.  

A less concentrated group of candidate protostars is associated with the well-studied W3(OH) subregion of massive star formation lying $\simeq 20$\arcmin\/ southeast of the W3~Main rich star cluster, including the very bright infrared source associated with X-ray selected candidate \#36.  With $[8.0] = 1.6$ mag, this is one of the most luminous protostars in the MCP catalog.  

Examination of Table~2 shows that the  infrared selected candidates \#35 and 37 in the W3(OH) subregion are also $Chandra$ X-ray sources that failed both of our X-ray selection criteria of $ME > 4.5$~keV and  $Cts < 10$.   

A few candidates are scattered about the field; these may represent distributed star formation in the cloud, or extragalactic contaminant populations that mimic X-ray emitting protostars.

\subsection{Flame Nebula}

The Flame Nebula (NGC~2024) is an H \textsc{ii} region and molecular cloud near the Orion Belt Star $\zeta$ Ori in the vicinity of the Horsehead Nebula within the large L1630 (Orion B) cloud. A dusty filament obscures the central star cluster in the optical and near-infrared bands.  The region has been well studied in the infrared \citep{Megeath12} and the X-ray \citep{Skinner03}. 

Most of the 23 MCPs lie in the main NGC~2024 cluster (subcluster `A') in an elongated north-south cloud filament.  Nine were previously identified as infrared protostars by \citet{Megeath12} and eight were previously noted as heavily absorbed X-ray sources by \citet{Skinner03}.  Note from column (12) of Table~2 that 10 of the MCPs are variable on short timescales in the X-ray band; this is a strong indicator they are young stars.  Candidate \#23 produces the 2~pc long outflow HH~92 \citep{Bally02}.  Candidate \#5 on the northern edge of the \textit{Herschel} core shows unusually high extinction ($ME \simeq$5.6~keV).  

Candidates \#1, 2, 4 and 20 lie $\sim 1$~pc to the south of the Flame Nebula cloud in the MM1/MM2 core region examined by \citet{Mookerjea09} using \textit{Spitzer} IRAC and MIPS and JCMT SCUBA. Candidate \#2 coincides with MIPS-5 classified as Class~0, and \#4 coincides with a lightly absorbed 2MASS star producing outflow HH~247.

\subsection{Rosette Nebula \label{Rosette.sec}}

The Rosette Nebula a mature giant H\textsc{ii} region ionized by the rich NGC 2244 cluster with more than 70 OB stars \citep{Roman08}.  The Rosette Molecular Cloud (RMC) extends $\sim 10$~pc to the east and southeast of the bubble, and molecular material is present in other directions as well.  The region has been extensively studied in both infrared and X-ray bands \citep[][and references therein]{Wang10, Ybarra13}.   Here we report a large sample of 67 candidate protostars.  All but a few candidates are selected by infrared excess; the ability to discover ultrahard X-ray sources is limited by the short (20~ks) $Chandra$ exposures in the RMC. 

A number of protostellar candidates (\#1, 2, 4, 6, 7, 8, 10, 11, 13, 22, 23) lie on the northern periphery of the ionizing NGC~2244 cluster in locations without prominent molecular material.  Active star formation in this area is unexpected.  They are unlikely to be members of the central cluster that is comprised mostly of old Class~III stars and resides in an H\textsc{ii} region evacuated of molecular material.  

A dense concentration of several candidates (\#15, 16, 17, 19, 20 and 21) appears in a poorly studied small molecular cloud $\sim 30\arcmin$\/ south of the Rosette Nebula outside the $Chandra$ field of view.  These are luminous protostars, and candidate \#16 appears to be associated with a shocked outflow. 

Candidates \#25 and 29 are massive embedded protostars lying on the western edge of the RMC near the rim of the H\textsc{ii} region.  Candidate \#29 = RMCX 89 = IRAS 06306+0437, luminous in both infrared and X-ray bands, is discussed by \citet{Wang09}. Candidate \#25, IRAS 06305+0440, is an infrared-only candidate protostar 2\arcmin\/ to the northwest, and Wang et al.\ suggest two additional Class~I protostars in the vicinity.  When Class~II and III stars are included, two small subclusters are found around these stars \citep{Kuhn14}.  The location and extreme youth of these subclusters suggest that they represent star formation triggered by the expanding H\textsc{ii} region.

About 20 protostellar candidates lie along a northwest-southeast ridge inside the RMC.  These are associated with near-infrared clusters PL~4, PL~5, PL~6 and REFL~8.  

We do not recover the high mass protostars AFGL961A, B and C reported by \citep{Williams09} in the embedded AFGL~961 cluster to the southeast. Infrared discovery is inhibited by bright nebular emission, and X-ray discovery is inhibited by the short $Chandra$ exposure. 

Candidate \#67 far to the northeast of the cloud is a $V \simeq 6$ B8~II giant star HR~2413 that is unabsorbed in the visible band but exhibits an infrared excess.  It is clearly not a protostar, but may be a member of an OB group spread over $\sim 4^\circ$ at a distance of 1.26~kpc, members of the larger Mon~OB2 association \citep{Kaltcheva11}.

\subsection{NGC 2264}

NGC`2264 is an extended, nearby star forming region in the northern sky, important in the historical understanding of pre-main sequence evolution.  Previous $Spitzer$ and $Chandra$ surveys have been conducted by \citet{Sung09}  and \citet{Flaccomio06}, respectively.  

Most of the 74 MYStIX protostellar candidates in NGC~2264 lie in two well-studied embedded clusters where there is little PAH emission to inhibit infrared photometry.  The sample is particularly rich in [4.5] excess emission infrared sources.  The molecular cloud hosting the northern rich `Spokes' star cluster, centered on IRS-1, has several dozen candidates. In a number of cases, the protostellar classification is validated by $Spitzer$ spectroscopy \citep{Forbrich10}.  Several lie in the concentrated microcluster of Class~0 submillimeter protostars in core D discovered by \citet{Teixeira07}.   Candidate \#44 is IRS~1 itself, an intermediate-mass B-type protostar discovered in the early years of infrared astronomy \citep{Allen72}\footnote{Our MCP sample does not include EXS-1 lying $\simeq 10$\arcsec\/ southwest of IRS~1.  This is the brightest x-ray source in the region found with the $XMM/Newton$ satellite \citep{Simon05} and is associated with a heavily obscured 2MASS star.  They report a powerful X-ray flare with absorption $\log N_H \simeq 22.9$ cm$^{-2}$ suggesting a Class~I protostar. However, the $Chandra$ spectrum shows $\log N_H \simeq 22.4$ cm$^{-2}$ with median energy $ME \simeq 2.7$~keV, considerably below the threshold of 4.5 keV for our protostellar candidates. This could be an example of variable X-ray absorption column densities occasionally seen in accreting Class~II stars such as AA~Tau and RW~Aur \citep{Grosso07, Schneider16}.}.   About 20 MCPs lie in the southern cluster centered on IRS~2. 

A handful of protostellar candidates are dispersed from the two main embedded clusters: 
\begin{enumerate}

\item Towards the south, candidates \#24 and 40 are near the Cone Nebula. The former emits in radio continuum and is aligned with visible outflow HH 125/225/226.  

\item Four MCPs appear in the northeast region around the O star S~Mon.   One of these, candidate \#47, is the famous Class~0 protostar NGC 2264G-VLA2 = IRAS 06384+0958 lying in a small, dense, isolated molecular core.  This is one of the original sources that defined the class \citep{Andre00}.  The presence of a group of protostars around S~Mon is unexpected as the majority of MYStIX stars in area are widely dispersed and less absorbed with a median age around 3~Myr using the X-ray/infrared photometric $Age_{JX}$ chronometer \citep{Getman14}.  This suggests that star formation is episodic in the S~Mon vicinity.  

\item Only one candidate protostar (\#2) is found around the Christmas Tree Nebula H\textsc{ii} region to the northwest.  

\end{enumerate}

\subsection{RCW 38}

This H\textsc{ii} region has very bright nebular emission and consequently its stellar population is difficult to study in the infrared.  The cluster is remarkable in having the highest stellar density of any cluster in the MYStIX survey \citep{Kuhn14}. An X-ray source list is provided by \citet{Wolk06} and, based on the X-ray sources, a $Spitzer$ study was performed by \citet{Winston11}.  They classified ten MCP stars as Class 0/I protostars by and several others are classified as flat spectrum YSOs.  Some have high infrared luminosities and are likely among the most massive members of the ionizing cluster.  
  
Candidate \#23 has extraordinary X-ray properties.  A single flare produces most of the detected 243 counts; with extremely heavy absorption ($ME \simeq 5.3$~keV), the inferred intrinsic peak X-ray luminosities is among the highest every seen in a stellar flare, similar to the X-ray flare recently reported from protostar ISO-Oph~85 in the $\rho$ Ophiuchi cloud \citep{Pizzocaro16}.   Several other candidates in RCW~38 showed X-ray flares though with $\sim$30 times lower luminosity than seen in \#23. 

A remarkably rich group of candidate protostars is associated with the cloud IRAS~09002-4732 around $02^h00^m$  $-45^\circ45^\prime$ lying $0.6^\circ$ southeast of the RCW~38 cluster.  These include MCP candidates \#56 through \#80 (except \#64), several of that are highly luminous massive protostars.  IRAS~09002-4732 is a small star formation region with $\sim 60$ Class~II stars dominated by an O9 star ionizing an ultracompact HII region G268.42-0.85 \citep{Apai05}.  We establish here the presence of very active star formation in the dusty filament oriented northwest-to-southeast that bifurcates the older star cluster.  Morphologically, the region resembles W~40 and the Flame Nebula.  

A dozen MCP sources are also found in small clouds scattered in all directions around RCW~38.  Candidates \#2 to the west and 19 to the south exhibit [4.5] excess characteristic of outflows, \#41 to the north was noted as an IRAS source, and \#3 to the southwest is particularly luminous.  Several ultrahard X-ray candidates are found, particularly in the very center of the cluster where PAH nebulosity inhibits infrared detection.  

\subsection{RCW 36}

The stellar population of this southern H\textsc{ii} region has not been studied before this work.  We find an ultrahard X-ray source and two infrared excess sources (one with [4.5] excess) associated with the small elongated molecular cloud.  Two other ultrahard X-ray sources lie several parsecs to the north; these could represent distributed star formation or extragalactic contamination.

\subsection{NGC 6334}

NGC 6334, nicknamed the Cat's Paw Nebula due to several lobate H\textsc{ii} regions fueled by a OB stars in many clusters of different ages, is a giant star-forming region on the Sagittarius-Carina spiral arm \citep{Willis13}.  Nearly 1700 MPCM stars are identified from the MYStIX survey \citep{Broos13} of which 57 are classified here as protostellar candidates.  The remaining 89 MCPs are MIRES sources satisfying our selection criteria found outside the $Chandra$ fields of view.  The spatial distribution is distinctive:  most candidates lie in or near the main $\sim 20$~pc northeast-to-southwest molecular filament, while the remaining are scattered in cloudlets at the ragged ends of the giant filamentary cloud.  This distribution was also found by Willis et al. (their Figure 11).  Where the H\textsc{ii} region emission is not too bright, the candidates emerge from the MIRES catalog, but in the central region the candidates are ultrahard X-ray sources.  Only a handful of the 146 MCPs lie in cloud-free regions.  

Except for the vicinity of the embedded massive protostars in NGC~6334~I(N), the rich stellar population of NGC~6334 has not be studied in detail due to the bright PAH infrared contamination, and consequently few of our protostellar candidates have been previously studied.  In particular, there is no catalog of OB stars responsible for ionizing the nebula.  Infrared excess surveys were performed by \citet{Robitaille08} and \citet{Willis13} and the X-ray source distribution was discussed by \citet{Feigelson09}.   

Candidates \#23, 66, 99, 116, 129   and 132 have particularly high bolometric luminosity with $5 < [8.0] < 7$ mag; two of these are associated with OH masers or [4.5] excess outflow emission.  Many other MCPs are consistent with intermediate-mass protostars.  Candidate \#49 is an ultrahard X-ray source at the center of the H\textsc{ii} region GPSR 351.248+0.667, and \#90 is associated with a faint millimeter source near NGC~6334~I(N).  Candidate \#51 is a well-studied interloper, an extragalactic blazar with strong radio, X-ray and gamma-ray emission that is projected by chance behind a star forming molecular core.  

\subsection{NGC 6357}

NGC~6357 is a giant H\textsc{ii} region illuminated by the lightly-obscured rich OB association Pismis~24 that contains several very massive O3-O2 stars.  It lies close to the star-forming region NGC~6334; the structures may have been formed in the same giant molecular cloud complex \citep{Russeil10}.   The region consists of the optical H\textsc{ii} region at the northern edge of a small interstellar bubble and two additional highly absorbed clusters in the molecular cloud to the east and southeast that were discovered in $Chandra$ images \citep{Townsley14}.  Except for the massive members of Pismis~24, the stellar population has not been extensively studied; an X-ray survey was performed by \citet{Wang07}, an infrared survey by \citet{Fang12}, and a visible survey by Russeil et al.  As a consequence, few of the 84 MYStIX candidate protostars presented  here have been previously reported.  

The spatial distribution of protostellar candidates is dramatically different from the older Class~II and III stars.  The older stars are concentrated into three rich clusters \citep{Broos13}, one revealed and two obscured, while the protostellar candidates are dispersed throughout the region.  The majority are associated with small  $Herschel$ clouds and filaments, but others are distributed widely to the south and east of the rich clusters. Several that appear in bright-rimmed pillars around the NGC~6357 bubble may represent star formation triggered by the expanding H\textsc{ii} region.  Nine MCPs brighter than $[8.0] = 7$~mag, and \#73 with maser emission, are likely massive protostars.  Candidate \#16, IRAS 17213-3437, is very luminous with $[8.0]=4.5$.  Several MCPs exhibit [4.5] excess emission. Seventeen candidates are ultrahard X-ray sources; most are associated with molecular cores though others are scattered and may be extragalactic contaminants.  

Altogether, the presence of several dozen MYStIX protostellar candidates, together with the collection of Class~I/flat spectrum sources found by \citet{Fang12}, indicates that NGC~6357 complex is actively forming stars today.

\subsection{Trifid Nebula} 

The Trifid Nebula is an optically bright emission nebula ionized by the O7 star HD~164492 that is trisected by three dust lanes. $Chandra$ X-ray and $Spitzer$ infrared studies of the region are presented by \citet{Rho04} and \citet{Rho06}.  We report 54 MCPs; 8 of these are detected as ultrahard X-ray sources and 12 are detected as [4.5] excess sources, and the remaining are detected by infrared color selection.  A dozen MCPs are concentrated in the molecular core adjacent to the central star cluster, while others are associated with cloud cores several parsecs south, north and east of the $Chandra$ field.  The majority of the infrared-selected MCPs are located at the end of an InfraRed Dark Cloud extending southward from the south corner of the $Chandra$ field of view. This IRDC has many MIRES sources that trace the filament \citep[Figure 9 of][]{Povich13}.

A few of the ultrahard X-ray sources may be extragalactic contaminants.   Nine MCPs were previously reported as protostellar candidates by Rho and colleagues.

\subsection{Lagoon Nebula}

The Lagoon Nebula (= Messier~8) has been long known as a star forming region with a prominent H\textsc{ii} region and a rich young cluster, NGC~6530 lying $<1^\circ$ from the Trifid Nebula. The main H\textsc{ii} region is fueled by several O stars dominated by the O4 star 9 Sag while the nearby Hourglass Nebula, a secondary compact H\textsc{ii} region, is ionized by the O7 star Herschel 36. The clusters are surrounded by a giant molecular cloud extending eastward. Earlier $Chandra$ and $Spitzer$ studies are reported by \citet{Damiani04} and \citet{Kumar10}, respectively.  $Herschel$ coverage of this MYStIX region is incomplete in the eastern portion, but the molecular cloud is evident in the $Spitzer$ maps.    

Thirty MYStIX candidate protostars, half of the 62 total MCPs, are selected as ultrahard X-ray sources; several were noted to have hard spectra by \citet{Damiani04}.  They are distributed widely across the molecular cloud.  Ten of the infrared-selected MCPs were previously identified by \citet{Kumar10}.  The candidate protostars do not show the clumpy clustering pattern of the older Class~II/III stars that are segmented into eleven elliptical subclusters by \citet{Kuhn14}.

\subsection{M 17} \label{m17.sec}

Messier 17 (= Omega Nebula), the second brightest H\textsc{ii} region in the sky after Orion, has the richest and most complex distribution of candidate protostars in this study with 180 MCPs.  The nebula is ionized by the rich cluster NGC~6618 with two O4+O4 binaries and dozens of other O stars \citep{Broos07}.  Several active star forming subregions have been studied  including:  M17 SW; the clumpy cloud on the southwest edge of the H\textsc{ii} region with luminous infrared stars like M17 UC-1 and the Kleinman-Wright Object; the M17 North cloud cores several arcminutes north of the nebula; and M17 SWex, an extended region $\sim 0.5^\circ$ west of the nebula lying outside the $Chandra$ fields.  Star formation in the M~17 region is studied in detail by Povich et al.(2009), Povich \& Whitney (2010), and \citet{Povich16}.  M~17 is also noted for its bright diffuse X-ray emission from shocked O star winds; this is also seen in other MYStIX regions with lower surface brightness \citep{Townsley11}. 

The rich sample of candidate protostars in M~17 arises in part from the unusually deep $Chandra$ exposures of the central region and from the extended coverage of surrounding regions ($\sim 50$~pc extent) by the $Spitzer$ and $Herschel$ satellites.   In the central region around NGC~6618, bright nebular emission reduces sensitivity to protostellar candidates selected by infrared SEDs.  About 75 MCPs are selected as ultrahard X-ray sources in the clumpy M~17~SW cloud; they are spatially distinct from the populous NGC~6618 cluster with $\sim 2000$ X-ray emitting Class~II and III members. 

Some of ultrahard X-ray candidates were originally catalogued by \citet{Broos07} although that study was based on shorter $Chandra$ exposures with less spatial coverage than available here.     Several candidates (e.g., \#92, 95, 101, 109, 116, 136) are X-ray bright with inferred unabsorbed X-ray luminosities at the top of the protostellar X-ray luminosity function ($\log Lx \simeq 32$~erg s$^{-1}$).  Most of these X-ray selected candidates are located in or near dark cores or filaments in $Spitzer$ images, often with associated localized polycyclic aromatic hydrocarbon nebulosity.   A few ultrahard X-ray candidates lie in an evacuated region east and northeast of M~17; some of these may be contaminant quasars.  

We briefly highlight some of the more interesting infrared selected MCPs in the M~17 region:
\begin{description}

\item[\#1-54]  Several dozen infrared selected MCPs are dispersed $>30$~pc from the principal cluster, particularly in the M17~SWex region to the west that is rich in intermediate-mass young stars \citep{Povich10, Povich16}.  

\item[\#55, 57-60] These are members of a dense group of protostars located in a bright $Herschel$ core 0.5$^\circ$  southwest of the $Chandra$ field.  The cloud has $M \simeq 600-1100$~M$_\odot$ of gas with associated H$_2$O maser emission. Candidate \#60 with [8.0]=8.7 and [4.5] excess emission is IRAS 18164-1631 = EGO G014.63-0.58; models of its SED give estimates of bolometric luminosity L$_{bol} \simeq  2$~L$_\odot$, stellar mass $M \simeq 2$~M$_\odot$, and envelope accretion rate $\dot{M} = 2 \times 10^{-4}$~M$_\odot$ yr$^{-1}$ \citep{Povich10}.  Candidate \#59 lying 23\arcsec\/ away, is very similar. However, \#58 lying 90\arcsec\/ to the north is far more luminous with [8.0]=5.1.  

\item[\#59, 60, 69, 74, 79, 95, 132, 133, 138, 168] These were previously identified as infrared sources with Class~0/I protostellar spectral energy distributions \citep{Povich09}.

\item[\#61] This infrared-selected star with [4.5] excess emission appears to be the bright unreddened foreground star Tyc 6265-637-1 = 2MASS J18192066-1543163 with B=12.2 and K=10.7.  It is not clear why it has an ascending mid-infrared spectrum. 

\item[\#95] This is an X-ray selected candidate that is also a very bright infrared source but with insufficient photometry to satisfy our infrared selection criteria. Its SED is modeled as a heavily reddened Class~I protostar with  L$_{bol}$ $\approx$ 1000 - 25000 L$_\odot$ and an estimate mass M $\approx$ 6 - 16 M$_\odot$. 

\item[\#110] This ultrahard X-ray star is previously known as near-infrared source Anon 1 \citep{Nielbock01}.  It is located in an area of bright nebulosity on the edge of a small filamentary  in the vicinity of ultracompact H\textsc{ii} region M17 UC1 and H$_2$O maser emission, its SED is modeled to give $L_{bol} \simeq 600$~M$_\odot$.  The X-ray source is heavily absorbed with $N_H \simeq  3 \times 10^{23}$~cm$^{-2}$ and  strong intrinsic luminosity $\log L_x $~(2-8 keV)~$\simeq 31.7$~erg s$^{-1}$.  

\item[\#112, 118, 120, 121, 123] These sources are members of a compact subcluster associated with the M17 UC1 ultracompact H\textsc{ii} region ionized by IRS~5 with estimated spectral type B0, and its associated M17~SWn submillimeter cloud core.  Candidate \#112 is an ultrahard X-ray star with very bright (sub)millimeter with $\sim 1000$~Jy flux at 450~$\mu$m.  Candidates \#121 and \#123 lie within $1.5$\arcsec\/ of IRS~5 that produces a small infrared nebulae and H$_2$O maser emission.  

\item[\#146] This high luminosity protostar ([8.0]=5.1) is the brightest infrared source in M~17N, a well-known secondary star forming cloud several parsecs north of the central NGC~6618.  With associated H$_2$O maser emission, it is modeled as an $5-10$~M$_\odot$ accreting protostar.  

\item[\#168] This [4.5] emission source shows jet-like structures in $Spitzer$ continuum and $H_2$ line maps \citep{CarattioGaratti15}. Although the $Spitzer$ source is not very bright ([8.0]=9.4), when far-infrared and millimeter emission is considered, the protostellar model has $L_{bol} \sim 1300$~L$_\odot$ with estimated star mass $M \sim 8$~M$_\odot$.   

\end{description}

\subsection{Eagle Nebula} \label{eagle.sec}

The Eagle Nebula (= Messier 16) is a complex molecular cloud shaped by the H\textsc{ii} cavity produced by the central NGC~6611 cluster.  Our MCP sample finds 87 protostellar candidates, most previously unidentified.   None appear around the so-called `Pillars of Creation' famously imaged with the {\it Hubble Space Telescope} suggesting these are not active star forming cores. The region is previously surveyed with $Chandra$ by \citep{Linsky07}, and a joint $Chandra - Spitzer$ survey of the region is reported by \citet{Guarcello09}.  

All but 19 of the MCPs are selected by virtue of their infrared characteristics.  The brightest infrared MCP with [8.0]=4.1 is candidate \#12 lying isolated south of the NGC~6611 cluster.   Candidate \#56 is a high luminosity protostar lying in a high-extinction molecular core HEC G017.19+00.81 MM2 within a prominent curved molecular filament \citep{Rygl13}.  This star is listed as an ultrahard X-ray source, part of the heavily extincted but sparse stellar subcluster G \citep{Kuhn14}, as its infrared colors did not satisfy our selection criteria.   Several other MCP candidates are in the vicinity of this subcluster with others associated with subcluster E lying 3\arcmin\/ to the south.  Candidate \#13 is the brightest infrared candidate in the field with [8.0]=4.1, and several others have [8.0]$\simeq 5-6$.  

The most active star forming locale in the region is the small cloud core at 18$^h$19$^m$09$^s$-13$^\circ$36\arcmin30\arcsec\/, G017.19+00.81 MM1, and the nearby curved molecular filament discussed by \citet{Rygl13}.  Over 20 MCP candidates, selected variously from infrared colors, [4.5] excess and ultrahard X-ray emission, are concentrated in these dense structures.  A larger molecular core on the western edge of the H\textsc{ii} contains candidates \#3-5 and \#7-10.  X-ray selected candidate \#26 lies in a dust pillar, and [4.5] excess candidate \#37 lies in the dense molecular tip of Pillar V seen as a bright $Herschel$ core with associated water masers.

\subsection{W 40}

This nearby but obscured star forming region shows a dramatic difference in young star distributions at X-ray and infrared wavebands.  The $Chandra$ image is dominated by a moderately rich and compact spherical star cluster at the center of the $Chandra$ field \citep{Kuhn10} while the infrared and submillimeter images are dominated by protostars in larger elongated molecular filaments west and north of the cluster \citep{Maury11}\footnote{Our infrared-selected candidate protostars in W~40 do not correspond well to a similar study of $JHK$ and $Spitzer$ images by \citet{Mallick13}.  Only three Class~I candidates are found in common, and these have positional offsets and photometric disagreements.  For example, many of the candidates reported by Mallick et al. lie in a $\sim 10$\arcmin\/ line oriented southeast from our candidate \#50 that is offset eastward from the $Herschel$ filament.  Their field of study does not include the $Herschel$ filament west of the H\textsc{ii} region where we find most of the MYStIX candidates.}.  

We find here a rich sample of 100 MCPs but only two ultrahard X-ray and three infrared color selected MCPs are within the X-ray cluster boundaries.  The great majority of the MYStIX  protostellar candidates are infrared-selected stars outside of the $Chandra$ field of view closely bound to a $\sim 4$~pc long southeast-to-northwest molecular filament lying to the west. This is the young embedded Serpens South cluster discovered by \citep{Gutermuth08}\footnote{Table 2 footnotes do not show association with the protostars found by \citet{Gutermuth08} because an electronic listing of their objects is not available.} from $Spitzer$ imaging that lies within a complex of interconnected, infalling filamentary clouds  \citep{Kirk13, Fernandez14}.  The MCP objects are concentrated in only certain portions of the filamentary system, and many of the candidates have [4.5] band excesses.  Several additional infrared MCPs are also present on a curved molecular  filament northeast  and a small filament to the far west of of the central W~40 cluster.
 
Various W~40 and Serpens South MCPs have interesting properties.  Seven candidates are unusually bright with [8.0]$< 6$ mag:  \#1 to far southwest has 2.8~mag; \#5 has 4.9~mag; \#36-38 have 5.4, 4.3 and 3.1~mag respectively;  \#69 has 5.0~mag; and \#88 has 5.6~mag.  These are likely higher mass protostars.  Candidates \#37, 38, and 67 produce $H_2$ or CO outflows.  Candidate \#63 coincides with SerpS-MM24, a faint millimeter source interpreted as a Class~0 protostar \citep{Maury11}.  Candidate \#81 coincides with W40 VLA~2, a faint radio continuum source without infrared counterpart.

\subsection{DR 21} \label{dr21.sec}

DR 21 is a clumpy molecular filament about 3~pc in extent lying within the larger Cygnus X star forming complex. The molecular ridge has three prominent dusty cloud cores with molecular `spokes' from smaller infalling filaments  \citep{Rivilla14}.  Our protostar search of the MYStIX catalogs produces 101 MCPs, most found through infrared excess.  Note that there is no bright H\textsc{ii} region producing nebulosity to impede detection of faint infrared sources in $Spitzer$ images.  Most of the protostellar candidates follow closely the north-south molecular ridge while some are associated with secondary molecular filaments or cloudlets in the region and others have infrared colors consistent with contaminant starburst galaxies.

A number of individual sources are worth noting.  Candidate \#14 lies at the tip of a small filament with a CO outflow.  
Candidate \#33 and 55 coincide with Cyg~N40~MM1 and Cyg-S-N48~MM2, respectively, that are small dense cloud fragments.
Candidate \#44 is an ultrahard X-ray source that is also the luminous infrared source W~75S~FIR~1, a well-known massive embedded protostar with H$_2$O maser emission. Candidate \#56 is the well-studied massive embedded protostar DR~21~(OH) with [8.0]=6.1, $L_{bol} \simeq 1000$ L$_\odot$, high-velocity outflow, and maser emission.  Candidate \#57 is the very luminous protostar DR21 IRS1 with [8.0]=2.9 lying at the center of a small cometary H\textsc{ii} region in the principal molecular concentration of the DR~21 filament.  Candidate \#79 is known as DR~21~FIR~3(H$_2$O), one of several infrared sources in a small dusty core fragment.   Candidate \#92 is the extremely red object ERO~1 lying at the edge of a sinuous InfraRed Dark Cloud producing a bipolar outflow.

\subsection{Control field: NGC 2362} \label{NGC2362.sec}
 
NGC 2362  is an age $\sim$ 5 Myr cluster in the MYStIX sample dominated by a single O9~I supergiant without any molecular material immediately around the cluster. Sources with infrared excesses from Class~II disks are rare, and we expect to find no Class~I protostars in the field \citep{Feigelson13}. In the X-ray, we find one candidate (071824.67-245310.7) with a median energy of 4.6~keV and a high flux of 73 net counts.   It is consistent with a background Type~2 quasar.  There are no infrared sources satisfying our criteria for protostellar candidates.   No image of this field appears in Figure~1.

\section{Discussion}

\subsection{Summary} \label{summary.sec}

The results of this search for protostars in the MYStIX joint infrared/X-ray survey of 14 OB-dominated star forming regions are summarized in Table~1.  We identify 1,109 MYStIX Candidate Protostars (MCPs) of which about 700 are newly reported here.  The great majority (955 candidates or 86\% of the total) are closely associated with dense cloud cores or filaments seen on $Herschel$ satellite far-infrared maps.  Except for the tightening of X-ray selection criteria to reduce candidates lying far from cloud cores (\S\ref{ultrahardX.sec}), the MCP selection criteria were based on infrared and X-ray photometric properties with no reference to location in the field.  

About 1/5 of the MCPs were identified by virtue of their X-ray photometry (\S\ref{ultrahardX.sec}) while most were identified using infrared photometry.  X-ray selection is particularly needed in locations where bright PAH nebulosity impedes detection of protostellar disks in the infrared.  We discuss the novel selection of protostars by X-ray spectral hardness in \S\ref{Xray.sec} below.   Our infrared selection procedures differ from those used by other groups, using combinations of SED modeling, location in the IRAC color-color diagram, and [4.5] band excess emission (\S\ref{irexcess.sec}).  In addition, photometric values are derived from MYStIX pipelines designed to treat crowded and nebulous regions in UKIDSS/UKIRT near-infrared images  and $Spitzer$ mid-infrared images \citep{King13, Kuhn13b}.  We therefore do not expect that MYStIX protostellar lists  to be identical to other infrared photometric studies.   

One useful result of the maps of MCPs in Figure~1 is the identification of new small star forming clouds in the vicinity of well-known young stellar clusters.  In the Rosette Nebula, for example,we see a previously unrecognized star forming cloudlet $\sim 10$~pc southwest of the main molecular cloud, and several cloudlets with active star formation are seen around RCW~38.  In other cases,  candidate protostars are found in areas where only older pre-main sequence stars were known, such as around S~Mon in NGC~2264 or in the IRAS~09002-4732 cloud near RCW~38.   
 
The reliability of the candidate list, discussed in \S\ref{reliability.sec}, appears to be high in the sense that the fraction of non-protostellar interlopers (False Positives) is low.  The main sources of contamination are expected to be extragalactic active galactic nuclei in the X-ray images, and Galactic AGB red giants and extragalactic starburst galaxies in the infrared images.   All of these populations should give spatial distributions that either avoid the molecular clouds or are randomly distributed across the fields.  But Figure~1 shows strongly clumped distributions of MCP stars along cloud cores and filaments.  This is clear evidence for the high reliability of our sample.  Interpretation of the scattered candidates is unclear; they may be residual contaminant populations, or indications of distributed star formation in small cloudlets.  

The completeness of the candidate list, discussed in \S\ref{incompleteness.sec}, is low.  The high fraction of MCPs that are very bright in the IRAC images $-$ such as those with [8.0]$ \leq 8$~mag  in the Rosette, NGC~6334, NGC~6357, Lagoon and Eagle Nebulae $-$ clearly suggests that a large population of lower brightness protostars is missing from the sample. We attempt to estimate the  incompleteness \S\ref{incompleteness.sec} and infer that the full population of protostars is probably an order of magnitude larger than our sample; that is, many thousands of protostars are probably present compared to our sample of $\sim 1,000$ MCPs in the 14 MYStIX regions under study.   Due to this large and uncertain incompleteness, we emphasize that the MCP list should not be used for astrophysical calculations that assume complete samples; for example, for estimation of star formation rates and efficiencies in molecular clouds.

\subsection{Reliability of results} \label{reliability.sec}

Several lines of evidence give confidence that most of these candidates are real protostars:
\begin{enumerate}

\item About half of the MCPs satisfied a different set of constraints to be listed MYStIX Probable Complex Member (MPCM) catalog of \citet{Broos13} that consists mainly of Class~II and III pre-main sequence stars.  The MPCM catalog is specifically designed to reduce contamination by Galactic field stars and extragalactic sources.  But it was not designed to capture protostars efficiently, particularly in the X-ray band (\S\ref{Xsel.sec}).    

\item As mentioned above, 86\%  of the MCPs are spatially associated with dense cores or filaments in $Herschel$ 500$\mu$m maps.  In NGC~6334, for example, several dozen MCPs  are closely aligned with the principal molecular filament or outlying cores. In NGC~2264, NGC~6334, Eagle Nebula, Trifid Nebula, W~40 and M~17, most candidates are aligned with the molecular structure.  In the Rosette Nebula, NGC~6357, Trifid Nebula, M~17, Eagle Nebula, W~40 and elsewhere the candidate protostars are spatially distinct from the older cluster rich in  Class~II and III stars ionizing the giant H\textsc{ii} region that was historically the first indication of star formation.   

\item X-ray selected MCPs are sometimes more dispersed; some coincide with cores and filaments, but others are widely dispersed in the $Chandra$ fields.  Some of the dispersed X-ray selected MCPs are probably the expected contaminants from Type 2 active galactic nuclei (\S\ref{AGN.contam.sec}). 

\item It is possible that a few of the MCPs are Class~II stars with nearly edge-on protoplanetary disks that can temporarily exhibit the high absorption expected in Class~I stars. The best studied example of such a system is AA~Tau, and \citet{Morales11} estimate that $\gtrsim 5$\% of Class~II stars in the Orion Nebula Cluster exhibit brief photometric dips from orbiting disk material in the $Spitzer$ IRAC bands. Two cases of X-ray absorption from an edge-on disk producing an ultrahard spectrum are also known in the Orion Nebula Cluster \citep[COUP~241 and COUP~419;][]{Kastner05}. 

\item The MCP sample recovers $\sim 300$ protostars found by past studies.  The level of association depends on the historical coverage of each region in the infrared.  For example, many MYStIX candidates are associated with known protostars in the M~17 region where past research has been extensive, but few are previously known in NGC~6357 where past study is weak.  As our listing is very incomplete (\S\ref{incompleteness.sec}), we do not expect to recover all previously known protostars.  

\end{enumerate}

Taking these issues into consideration, we estimate that $80-90$\% of the 1,109 MCPs are likely to be true protostars.  The selection criteria derived in \S\ref{selection.sec} was sufficiently conservative to remove the large contaminating populations of dusty Galactic field stars and obscured extragalactic objects from our sample.

\subsection{Incompleteness of the protostar survey} \label{incompleteness.sec}

The protostar selection procedures described in \S\ref{selection.sec} are very conservative in the sense that criteria were chosen at levels to increase reliability of the sample (i.e., set to reduce false positives) and not chosen to capture a large fraction of the true protostellar population (i.e., set to reduce false negatives).  These constraints can be summarized as follows: 
\begin{enumerate}

\item Our criterion of $ME>4.5$~keV requires extremely high absorption to the X-ray source, and excludes the majority of protostars that have $3.0 < ME < 4.5$~keV ($22.5 <\log N_H < 23.3$~cm$^{-2}$) range.  See Table~1 of Imanishi et al.\ 2001, Tables~2 and 4 of Grosso et al.\ 2005 and Table~6 of Prisinzano et al.\ 2008 for X-ray hardness distributions of protostars in the $\rho$ Ophiuchi and Orion Nebula regions. A few of these less-absorbed X-ray emitting protostars are found through our infrared excess criteria; for example, MCP Flame 7 has 36 net counts with $ME = 3.6$~keV.   

\item The X-ray flux limits for heavily absorbed stars in the MYStIX fields severely truncates our sample.  For a hard intrinsic spectrum (say, thermal bremsstrahlung with $kT = 5$~keV or $T \simeq 100$~MK), less than 5\% of the photons that would be detected by the $Chandra$ ACIS detector from a hypothetical unabsorbed protostar are in fact detected when $ME > 4.5$~keV.  Thus our requirement $>10$ net counts is equivalent to a requirement of $>200$ net counts if absorption were not present.  This essentially confines the ultrahard X-ray sources to the upper tail of the protostellar X-ray luminosity function with $\log L_X \simeq 31-32$~erg s$^{-1}$.   Protostars with X-ray flares at this level have been found \citep{Hamaguchi05, Pizzocaro16} but are rare \citep{Imanishi01, McCleary11, Prisinzano08}.

\item The infrared color-color criteria should detect many Class~I protostars, limited by the impacts of crowding and spatially varying PAH nebular emission.  But the criterion of [4.5]$\mu$m excess emission excludes the large majority of infrared sources with protostar-like SEDs (see Figures~3 and 6 of Povich et al.\ 2013).  This photometric characteristic is sufficient indicator of protostars but is not at all a necessary indicator.

\item Most Class~0 protostars, and any Class~I stars with extremely heavy absorption ($A_V \geq 200$~mag), are missed by our procedures described in \S\ref{selection.sec}.  Possible Class~0 sources captured in the MCP catalog include SerpS-MM24 (MCP W40 63), W40 MM5 (MCP W40 79), and members of the Class~0 microcluster in NGC~2264 (MCP NGC2264 27, 30-32).  For extremely heavily absorbed stars, photometric identification requires longer wavelengths than available from $Spitzer$'s IRAC detector and higher sensitivity than available from $Chandra$'s X-ray telescope with typical exposure times. For example, a deep $Chandra$ observation of the Serpens cloud detected 9 Class~I and 9 Class~I/II (flat spectrum) protostars, but none of the 6 Class~0 or Class~0/I stars \citep{Giardino07}.  

\end{enumerate}

Combining these effects is not simple.  The survey depths in the X-ray and infrared bands differ considerably among regions, so quantitative conclusions about the full MYStIX survey are not reliable.  The relationship between sensitivity and protostellar mass is not clear for $any$ protostellar survey; completeness with respect to the underlying Initial Mass Function is thus always difficult to establish.   Our rough qualitative estimate based on comparison with previous protostellar surveys that our selection procedure in \S\ref{selection.sec} misses at least half, and probably most, of the true protostellar population.  The full underlying protostellar population would then be several our observed 1,109 MCPs in all 14 MYStIX regions examined.  

Another rough estimate of the incompleteness can be made with the mathematics of `capture-recapture' modeling used for over a century in ecological studies \citep{Amstrup05}.  Here we estimate the total population of a class based on the overlap when two independent but incomplete samples are obtained assuming equal probability of capture in each experiment.  The population is assumed to be 'closed' with no additions or losses among the experiments.  Consider the first 'capture' experiment as the detection of candidate protostars based on infrared excess.  These are then `returned' to the environment and a second `capture' experiment detects candidate protostars based on X-ray properties.   We omit X-ray observations in regions where strong PAH nebulosity inhibits infrared detection, and we omit infrared observations in regions without $Chandra$ detection.  These regions have only one effective `capture' experiment.

The fact that very few protostars are simultaneously detected with X-ray and infrared selection techniques is, in the context of capture-recapture theory, indicative that the total population is much larger than the combined samples.  We treat the infrared detection of $\sim 700$ MCPs in $Chandra$ fields to be the first `capture' experiment, and the X-ray detection of $\simeq 100$ candidate protostars in regions where PAH contamination is not severe to be the second `recapture' experiment.  In our MCP survey, only five sources are recovered by both X-ray and infrared methods: MCP Flame 14, MCP NGC2264 36, MCP NGC2264 54, MCP DR21 32, and MCP DR21 66.   However, about 70 candidates would have been recovered by both surveys is the source ha redder infrared colors, a few more X-ray photons, or higher X-ray median energy.    Standard mathematical procedures from capture-recapture analysis are used: the Petersen estimator of the total protostellar population with Chapman's bias correction and Seber's standard error estimate \citep{Amstrup05}.  This computation was made with function $closedp.bc$ in CRAN package $Rcapture$ \citep{Rivest14} within the public domain R statistical software environment \citep{RCoreTeam15}.

For 5 recovered stars from experiments with 700 and 100 captures, the estimated total population of $27 \pm 11$ thousand protostars. If we more optimistically consider 70 stars recovered using both methods by relaxing X-ray and infrared selection rules, the estimated population would be  $3.1 \pm 0.3$ thousand protostars.   We conclude that there are $\sim 3,000 - 30,000$ protostars in these 14 MYStIX fields, so that the MCP identification rate is between 3\% and 30\% of the full protostellar population.  


We conclude that the MCP sample provides a very incomplete complete survey of the full protostellar population in the 14 examined MYStIX fields.  An order of magnitude more protostars are probably present that we do not identify, and quantitative estimates of incompletness are very inaccurate.  {\it As a consequence of this large and poorly quantified incompleteness factor, the MYStIX Protostellar Candidates sample should not be used for any statistical or astrophysical purpose, such as estimation of the  current star formation rate or efficiency in these molecular clouds.}

\subsection{The nature of X-ray ultrahard sources} \label{Xray.sec}

Perhaps the most unusual element of our search for MYStIX protostars is the inclusion of ultrahard X-ray sources that do not have known infrared counterparts.  This may seem risky, as infrared excess with an ascending SED has historically been the sole criterion for identifying Class~I protostars.  But the risk may not be so high.  First, the MYStIX fields are inhospitable for self-consistent and sensitive identification of protostars due to sensitivity reduction from the nebulosity of bright H\textsc{ii} regions, contamination by dusty red giants, and crowding. So the absence of catalogued infrared sources from low-resolution telescopes like $Spitzer$ associated with ultrahard X-ray sources does not clearly indicate that a protostar is not present.  Second, the combination of X-ray selection criteria of $ME > 4.5$~keV and Net counts $>10$ is so severe that, except for a handful of randomly located Type~2 quasars, the resulting sample should be confined to X-ray emitting Class~I stars.  Third, most of the the ultrahard sources are clustered around $Herschel$ cloud cores and, in some cases, recover previously known protostars.  

We therefore encourage followup of the X-ray ultrahard MCP sources using infrared telescopes.  It is possible that they represent a subclass of protostars with different properties than those located with traditional infrared color criteria.  The most valuable instrumental characteristic for followup would be imaging and photometry at high spatial resolution at $2-10 \mu$m wavelengths, needed to reduce the surface brightness of nebulosity.  Large ground-based telescopes with active optics in the infrared band, and NASA's forthcoming {\it James Webb Space Telescope}, would be particularly effective for finding the expected infrared counterparts.

\acknowledgments
The first two authors contributed equally to this study.  We thank Pat Broos, Leisa Townsley (Penn State) and Tim Naylor (Exeter) for their crucial contributions to the MYStIX project. MYStIX has been supported at Penn State by NASA grant NNX09AC74G, NSF grant AST-0908038, and the $Chandra$/ACIS Team contract SV4-74018 issued by SAO/CXC under contract NAS8-03060.  M.S.P. is supported by the NSF through grant CAREER-1454333. This research has made use of NASA's Astrophysics Data System, the SIMBAD database operated at CDS (Strasbourg, France), and the NASA/IPAC Infrared Science Archive, operated by the JPL/Caltech under contract with NASA.  

{\it Facilities:} \facility{CXO (ACIS)}, \facility{Spitzer (IRAC)}, \facility{UKIRT (WFCAM)}

\newpage

\begin{figure*}[t]
  \centering
  \includegraphics[width=1.0\linewidth]{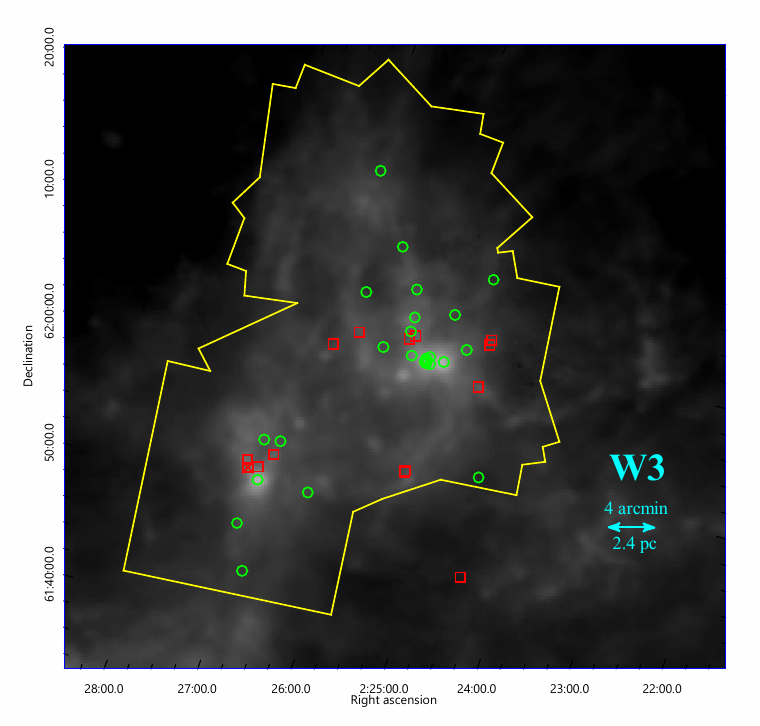}
\caption{MYStIX fields with candidate protostars.  The background gray-scale map is the 500~$\mu$m $Herschel$ SPIRE instrument shown on a logarithmic scale. The yellow outline shows the $Chandra$ X-ray field of view within which MYStIX Probable Complex Members are located.  Infrared excess stars are extracted from the MIRES catalog \citep{Povich13} and can lie outside the $Chandra$ field.  The candidate protostars from Table~2 are superposed:  infrared excess sources with [4.5] excess (red $\times$), other infrared excess sources (red box), and ultrahard X-ray sources (green circle).   Panels show regions: W~3, Flame Nebula, Rosette Nebula, NGC~2264, RCW~38, RCW~36, NGC~6334, NGC~6357, Trifid Nebula, Lagoon Nebula, M~17, Eagle Nebula, W~40, and DR~21. The full set of panels appears as a figure set in the electronic journal.}
\label{fig:MCP_figures}
\end{figure*}

%
%
%
%
%
%
%
%
%
%
%
%

\figsetstart
\figsetnum{1}
\figsettitle{MYStIX fields with candidate protostars}

\figsetgrpstart
\figsetgrpnum{1.1}
\figsetgrptitle{W 3}
\figsetplot{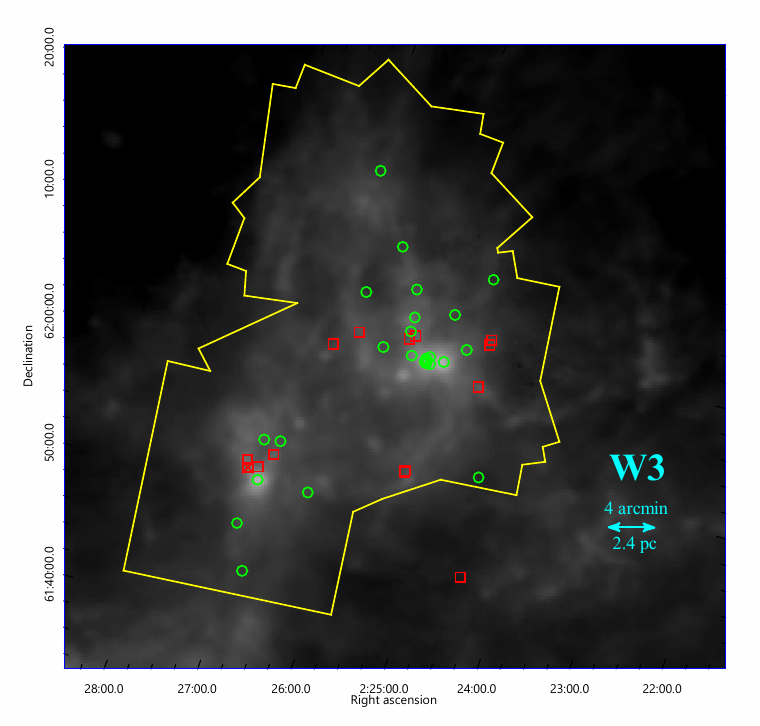}
\figsetgrpnote{W 3 MYStIX field with candidate protostars. The background gray-scale map is the 500 μm Herschel SPIRE instrument shown on a logarithmic scale. The yellow outline shows the Chandra X-ray field of view within which MYStIX Probable Complex Members are located. Infrared excess stars are extracted from the MIRES catalog (Povich et al. 2013) and can lie outside the Chandra field. The candidate protostars from Table 2 are superposed: infrared excess sources with [4.5] excess (red X), other infrared excess sources (red box), and ultrahard X-ray sources (green circle).}
\figsetgrpend

\figsetgrpstart
\figsetgrpnum{1.2}
\figsetgrptitle{Flame Nebula}
\figsetplot{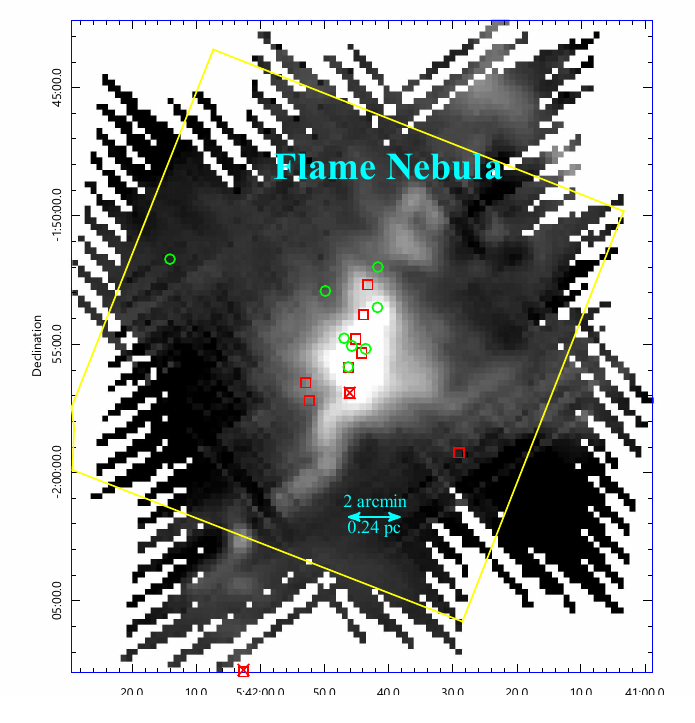}
\figsetgrpnote{Flame Nebula MYStIX field with candidate protostars. The background gray-scale map is the 500 μm Herschel SPIRE instrument shown on a logarithmic scale. The yellow outline shows the Chandra X-ray field of view within which MYStIX Probable Complex Members are located. Infrared excess stars are extracted from the MIRES catalog (Povich et al. 2013) and can lie outside the Chandra field. The candidate protostars from Table 2 are superposed: infrared excess sources with [4.5] excess (red X), other infrared excess sources (red box), and ultrahard X-ray sources (green circle).}
\figsetgrpend

\figsetgrpstart
\figsetgrpnum{1.3}
\figsetgrptitle{Rosette Nebula}
\figsetplot{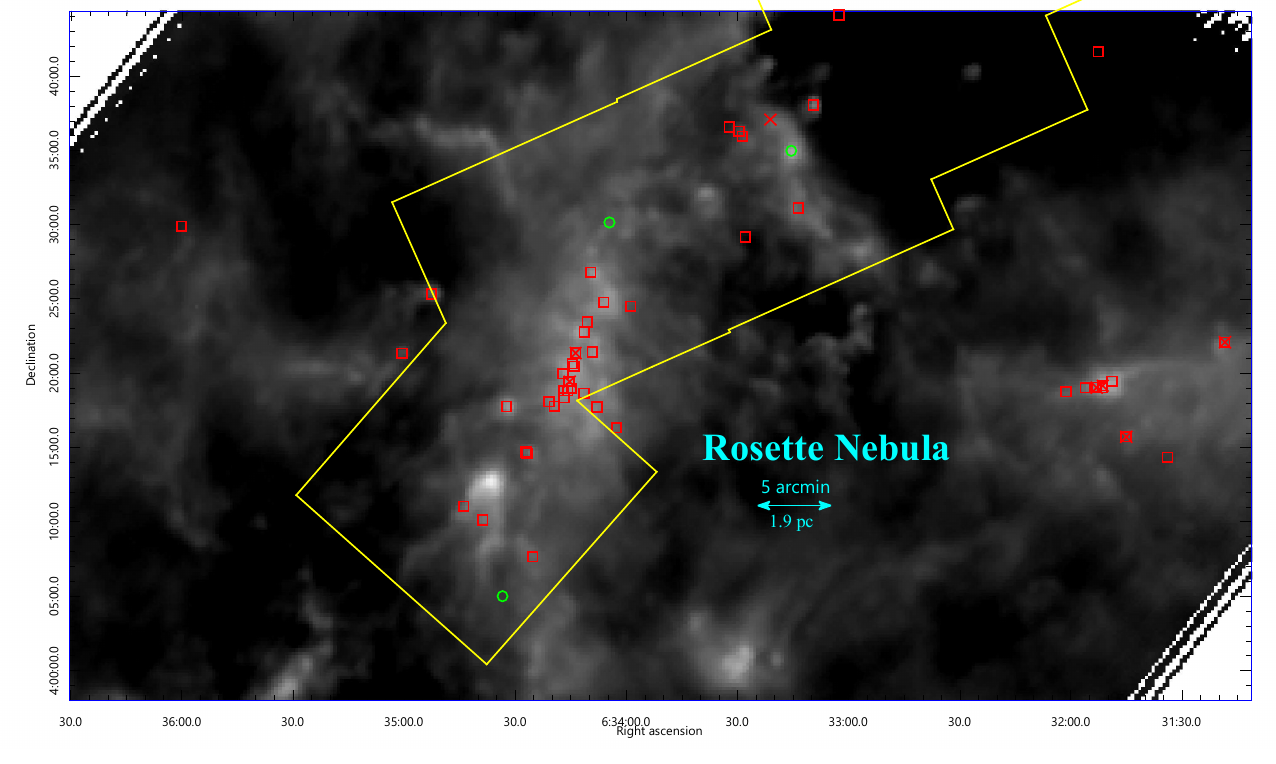}
\figsetgrpnote{Rosette Nebula MYStIX field with candidate protostars. The background gray-scale map is the 500 μm Herschel SPIRE instrument shown on a logarithmic scale. The yellow outline shows the Chandra X-ray field of view within which MYStIX Probable Complex Members are located. Infrared excess stars are extracted from the MIRES catalog (Povich et al. 2013) and can lie outside the Chandra field. The candidate protostars from Table 2 are superposed: infrared excess sources with [4.5] excess (red X), other infrared excess sources (red box), and ultrahard X-ray sources (green circle).}
\figsetgrpend

\figsetgrpstart
\figsetgrpnum{1.4}
\figsetgrptitle{NGC 2264}
\figsetplot{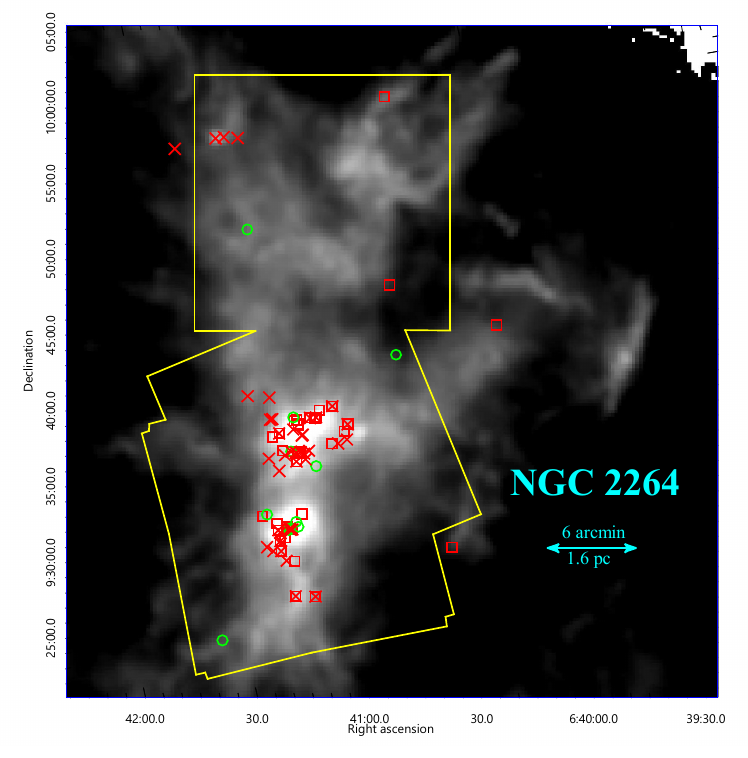}
\figsetgrpnote{NGC 2264 MYStIX field with candidate protostars. The background gray-scale map is the 500 μm Herschel SPIRE instrument shown on a logarithmic scale. The yellow outline shows the Chandra X-ray field of view within which MYStIX Probable Complex Members are located. Infrared excess stars are extracted from the MIRES catalog (Povich et al. 2013) and can lie outside the Chandra field. The candidate protostars from Table 2 are superposed: infrared excess sources with [4.5] excess (red X), other infrared excess sources (red box), and ultrahard X-ray sources (green circle).}
\figsetgrpend

\figsetgrpstart
\figsetgrpnum{1.5}
\figsetgrptitle{RCW 38}
\figsetplot{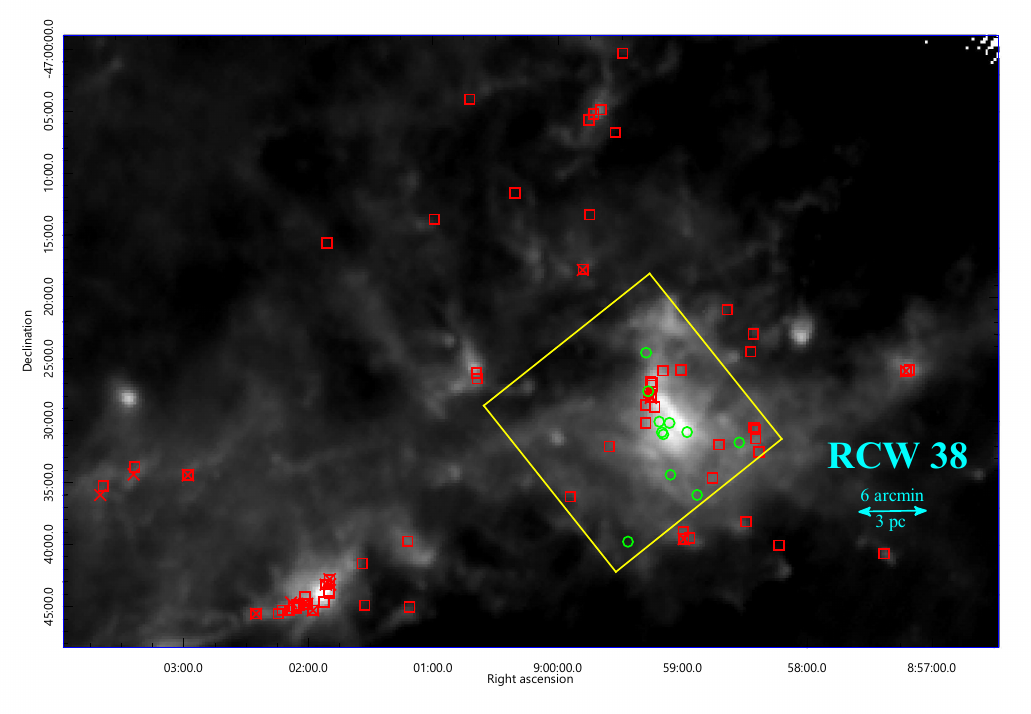}
\figsetgrpnote{RCW 38 MYStIX field with candidate protostars. The background gray-scale map is the 500 μm Herschel SPIRE instrument shown on a logarithmic scale. The yellow outline shows the Chandra X-ray field of view within which MYStIX Probable Complex Members are located. Infrared excess stars are extracted from the MIRES catalog (Povich et al. 2013) and can lie outside the Chandra field. The candidate protostars from Table 2 are superposed: infrared excess sources with [4.5] excess (red X), other infrared excess sources (red box), and ultrahard X-ray sources (green circle).}
\figsetgrpend

\figsetgrpstart
\figsetgrpnum{1.6}
\figsetgrptitle{RCW 36}
\figsetplot{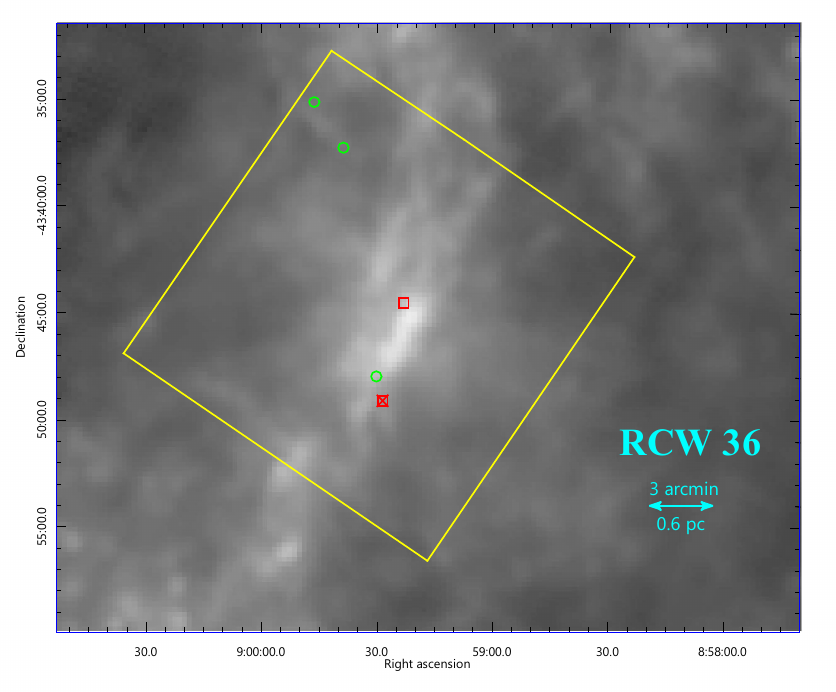}
\figsetgrpnote{RCW 36 MYStIX field with candidate protostars. The background gray-scale map is the 500 μm Herschel SPIRE instrument shown on a logarithmic scale. The yellow outline shows the Chandra X-ray field of view within which MYStIX Probable Complex Members are located. Infrared excess stars are extracted from the MIRES catalog (Povich et al. 2013) and can lie outside the Chandra field. The candidate protostars from Table 2 are superposed: infrared excess sources with [4.5] excess (red X), other infrared excess sources (red box), and ultrahard X-ray sources (green circle).}
\figsetgrpend

\figsetgrpstart
\figsetgrpnum{1.7}
\figsetgrptitle{NGC 6334}
\figsetplot{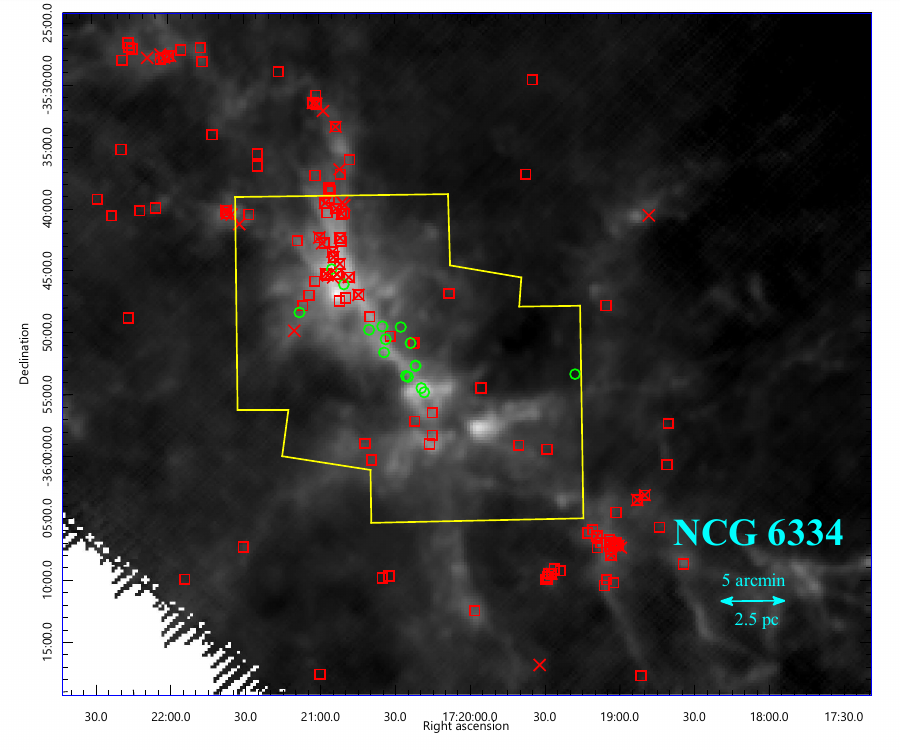}
\figsetgrpnote{NGC 6334 MYStIX field with candidate protostars. The background gray-scale map is the 500 μm Herschel SPIRE instrument shown on a logarithmic scale. The yellow outline shows the Chandra X-ray field of view within which MYStIX Probable Complex Members are located. Infrared excess stars are extracted from the MIRES catalog (Povich et al. 2013) and can lie outside the Chandra field. The candidate protostars from Table 2 are superposed: infrared excess sources with [4.5] excess (red X), other infrared excess sources (red box), and ultrahard X-ray sources (green circle).}
\figsetgrpend

\figsetgrpstart
\figsetgrpnum{1.8}
\figsetgrptitle{NGC 6357}
\figsetplot{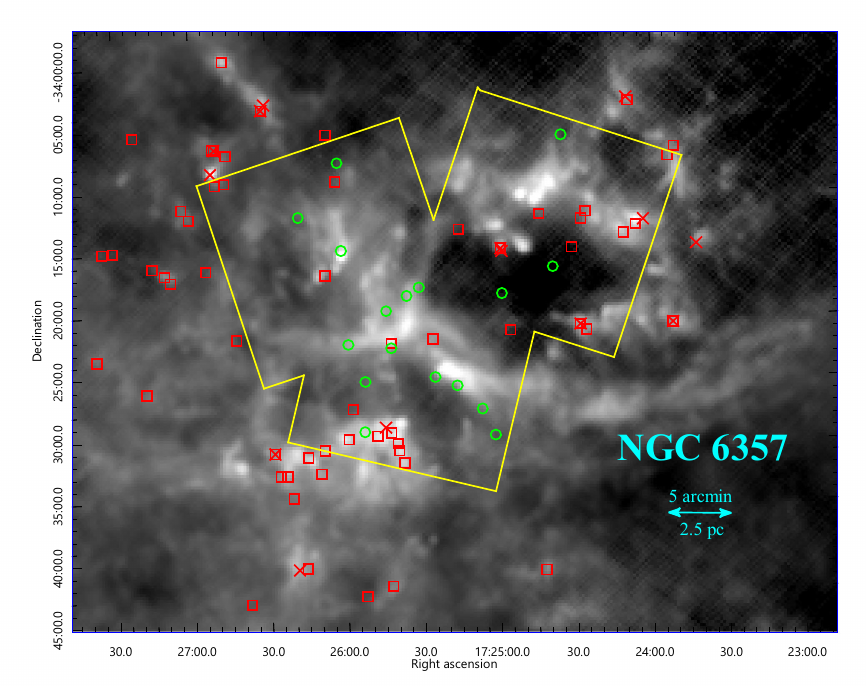}
\figsetgrpnote{NGC 6357 MYStIX field with candidate protostars. The background gray-scale map is the 500 μm Herschel SPIRE instrument shown on a logarithmic scale. The yellow outline shows the Chandra X-ray field of view within which MYStIX Probable Complex Members are located. Infrared excess stars are extracted from the MIRES catalog (Povich et al. 2013) and can lie outside the Chandra field. The candidate protostars from Table 2 are superposed: infrared excess sources with [4.5] excess (red X), other infrared excess sources (red box), and ultrahard X-ray sources (green circle).}
\figsetgrpend

\figsetgrpstart
\figsetgrpnum{1.9}
\figsetgrptitle{Trifid Nebula}
\figsetplot{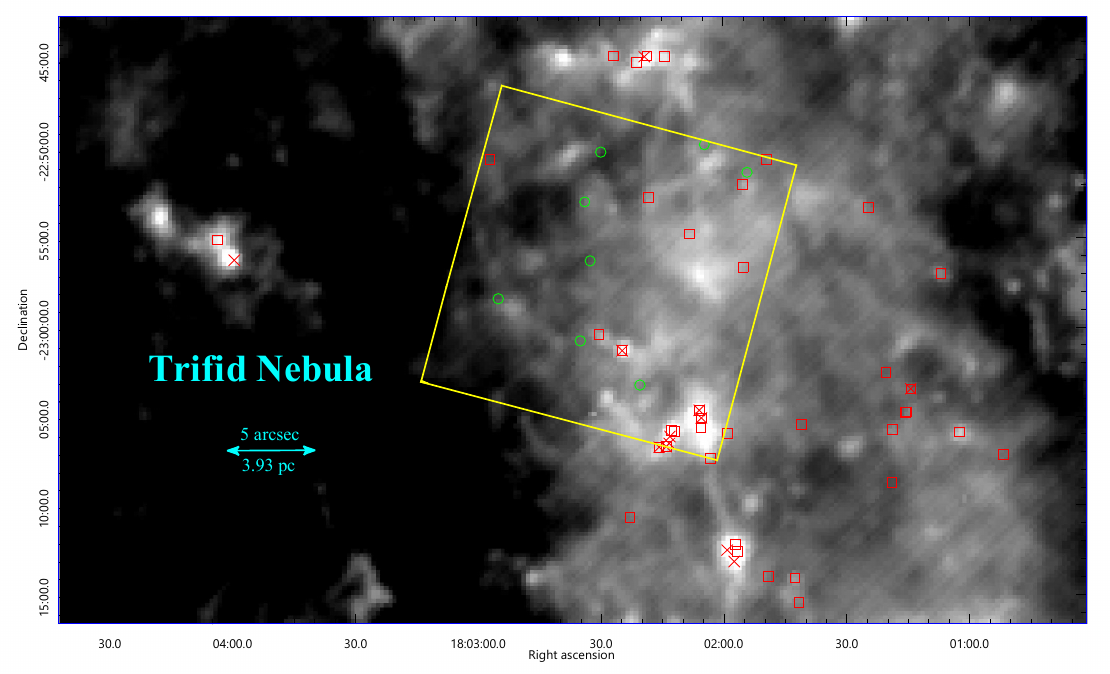}
\figsetgrpnote{Trifid Nebula MYStIX field with candidate protostars. The background gray-scale map is the 500 μm Herschel SPIRE instrument shown on a logarithmic scale. The yellow outline shows the Chandra X-ray field of view within which MYStIX Probable Complex Members are located. Infrared excess stars are extracted from the MIRES catalog (Povich et al. 2013) and can lie outside the Chandra field. The candidate protostars from Table 2 are superposed: infrared excess sources with [4.5] excess (red X), other infrared excess sources (red box), and ultrahard X-ray sources (green circle).}
\figsetgrpend

\figsetgrpstart
\figsetgrpnum{1.10}
\figsetgrptitle{Lagoon Nebula}
\figsetplot{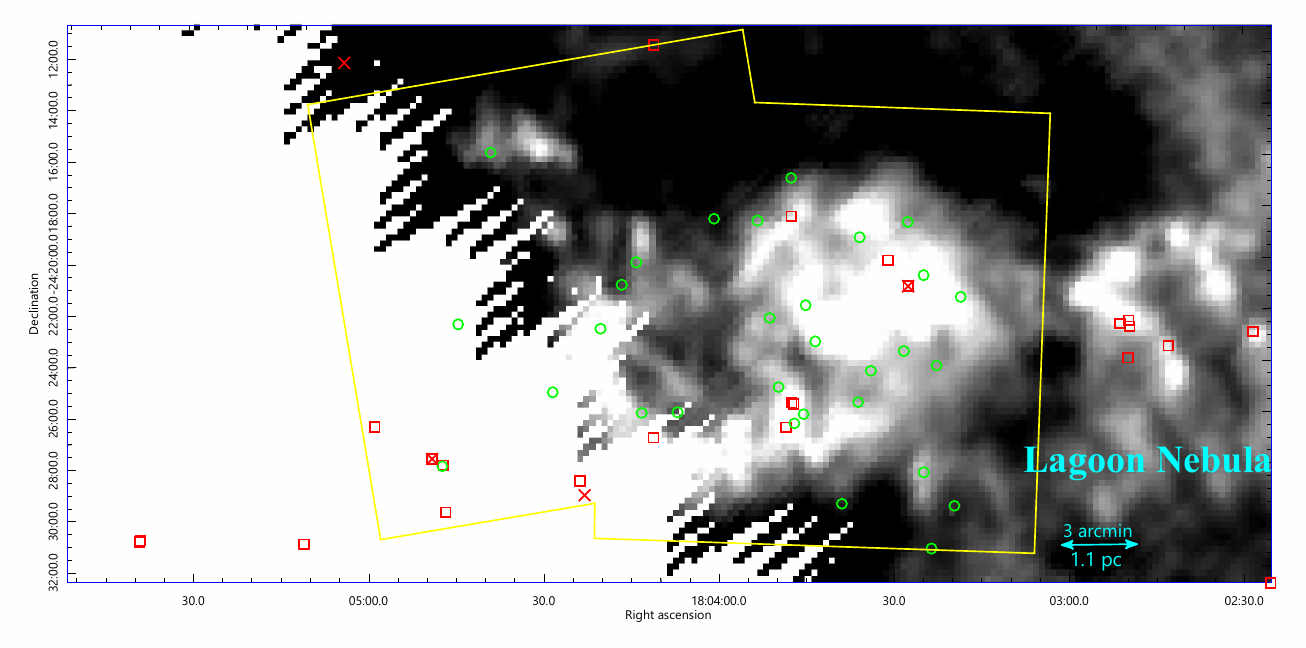}
\figsetgrpnote{Lagoon Nebula MYStIX field with candidate protostars. The background gray-scale map is the 500 μm Herschel SPIRE instrument shown on a logarithmic scale. The yellow outline shows the Chandra X-ray field of view within which MYStIX Probable Complex Members are located. Infrared excess stars are extracted from the MIRES catalog (Povich et al. 2013) and can lie outside the Chandra field. The candidate protostars from Table 2 are superposed: infrared excess sources with [4.5] excess (red X), other infrared excess sources (red box), and ultrahard X-ray sources (green circle).}
\figsetgrpend

\figsetgrpstart
\figsetgrpnum{1.11}
\figsetgrptitle{M 17}
\figsetplot{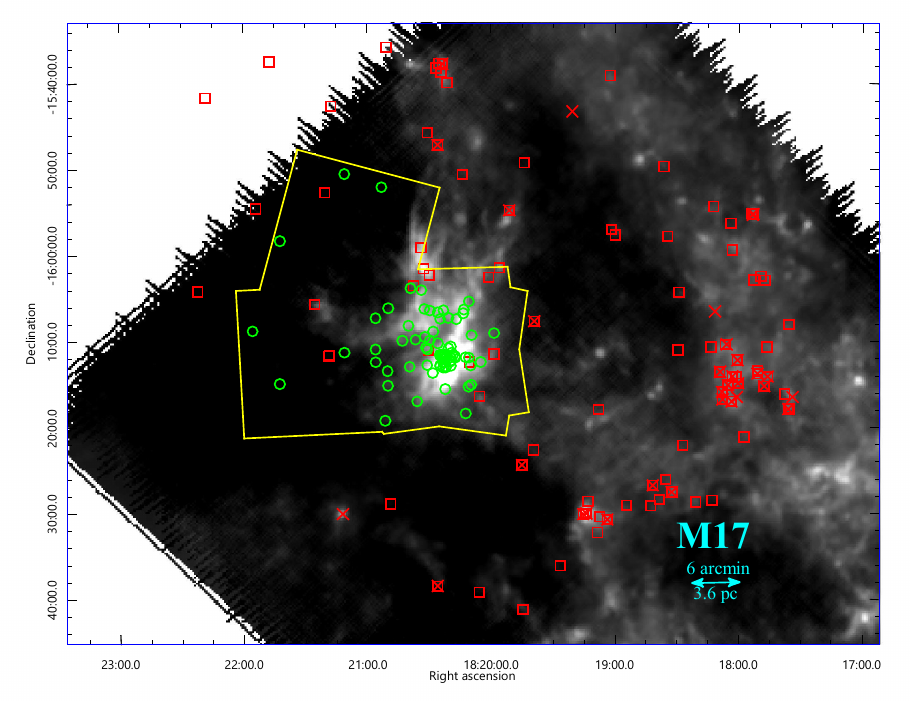}
\figsetgrpnote{M 17 MYStIX field with candidate protostars. The background gray-scale map is the 500 μm Herschel SPIRE instrument shown on a logarithmic scale. The yellow outline shows the Chandra X-ray field of view within which MYStIX Probable Complex Members are located. Infrared excess stars are extracted from the MIRES catalog (Povich et al. 2013) and can lie outside the Chandra field. The candidate protostars from Table 2 are superposed: infrared excess sources with [4.5] excess (red X), other infrared excess sources (red box), and ultrahard X-ray sources (green circle).}
\figsetgrpend

\figsetgrpstart
\figsetgrpnum{1.12}
\figsetgrptitle{Eagle Nebula}
\figsetplot{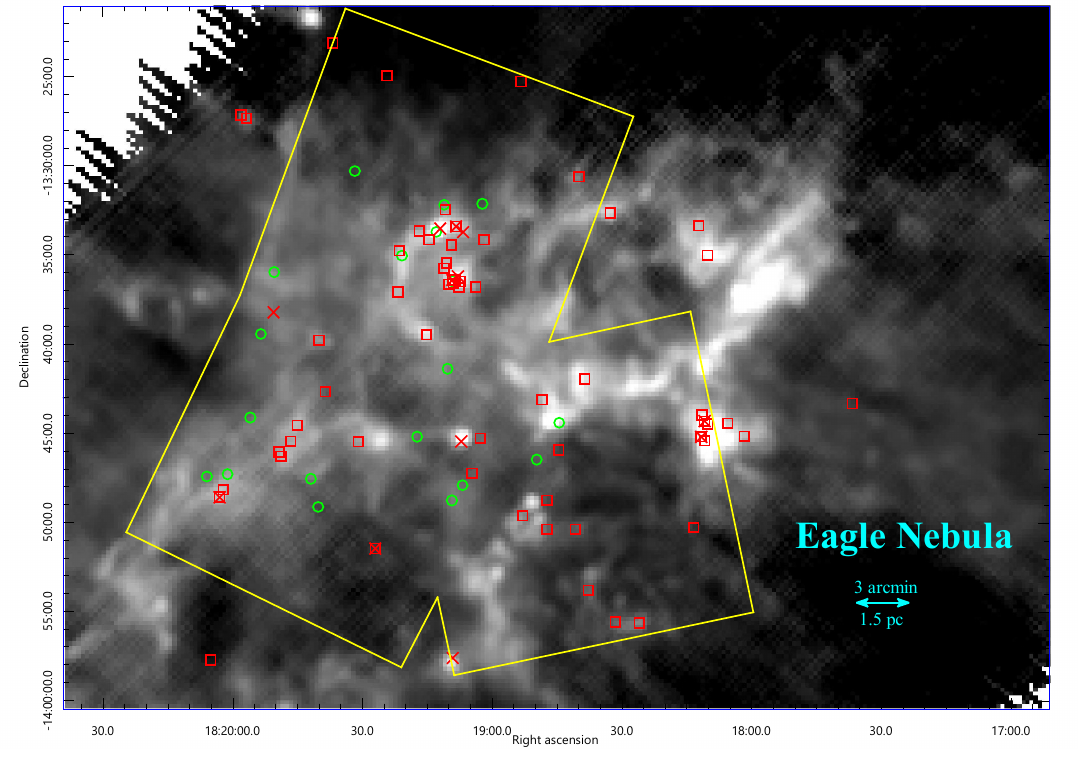}
\figsetgrpnote{Eagle Nebula MYStIX field with candidate protostars. The background gray-scale map is the 500 μm Herschel SPIRE instrument shown on a logarithmic scale. The yellow outline shows the Chandra X-ray field of view within which MYStIX Probable Complex Members are located. Infrared excess stars are extracted from the MIRES catalog (Povich et al. 2013) and can lie outside the Chandra field. The candidate protostars from Table 2 are superposed: infrared excess sources with [4.5] excess (red X), other infrared excess sources (red box), and ultrahard X-ray sources (green circle).}
\figsetgrpend

\figsetgrpstart
\figsetgrpnum{1.13}
\figsetgrptitle{W 40}
\figsetplot{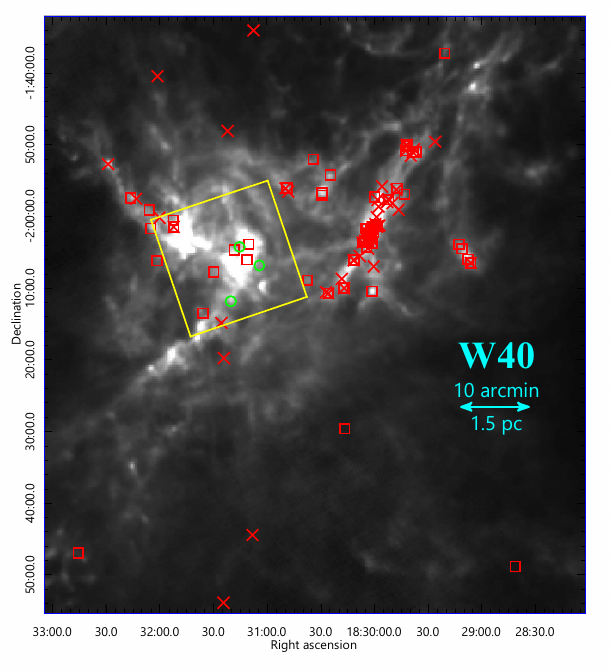}
\figsetgrpnote{W 40 MYStIX field with candidate protostars. The background gray-scale map is the 500 μm Herschel SPIRE instrument shown on a logarithmic scale. The yellow outline shows the Chandra X-ray field of view within which MYStIX Probable Complex Members are located. Infrared excess stars are extracted from the MIRES catalog (Povich et al. 2013) and can lie outside the Chandra field. The candidate protostars from Table 2 are superposed: infrared excess sources with [4.5] excess (red X), other infrared excess sources (red box), and ultrahard X-ray sources (green circle).}
\figsetgrpend

\figsetgrpstart
\figsetgrpnum{1.14}
\figsetgrptitle{DR 21}
\figsetplot{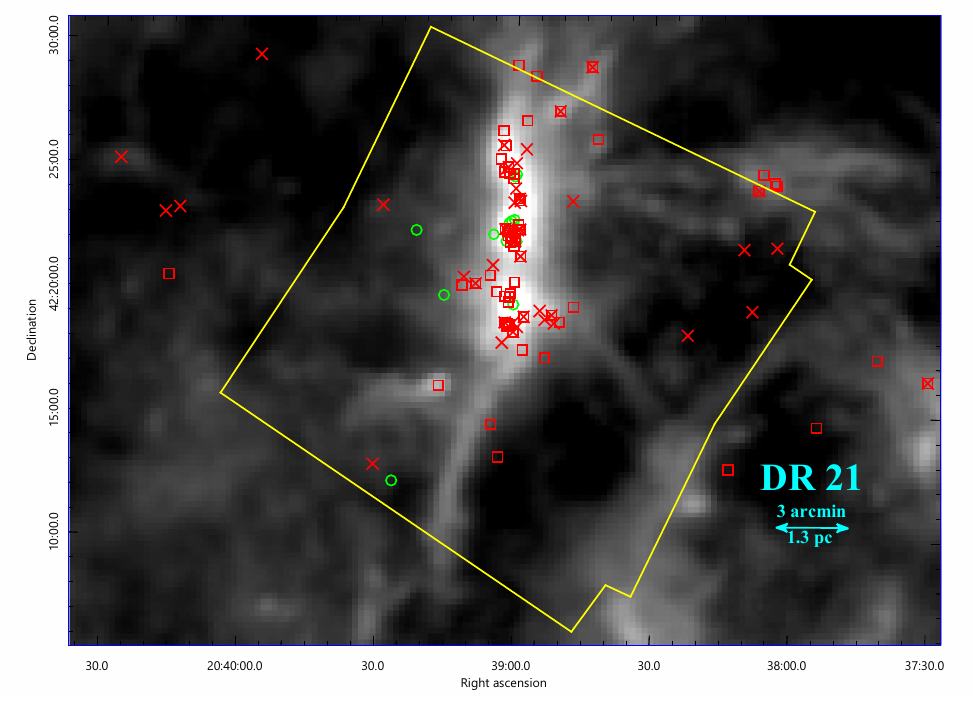}
\figsetgrpnote{DR 21 MYStIX field with candidate protostars. The background gray-scale map is the 500 μm Herschel SPIRE instrument shown on a logarithmic scale. The yellow outline shows the Chandra X-ray field of view within which MYStIX Probable Complex Members are located. Infrared excess stars are extracted from the MIRES catalog (Povich et al. 2013) and can lie outside the Chandra field. The candidate protostars from Table 2 are superposed: infrared excess sources with [4.5] excess (red X), other infrared excess sources (red box), and ultrahard X-ray sources (green circle).}
\figsetgrpend

\figsetend

\newpage


\begin{deluxetable}{lrrrrrrrr}
\setlength{\tabcolsep}{0.05in} 
\tablecaption{Samples of Protostellar Candidates \label{summary.tbl}}
\tablewidth{0pt}
\tablehead{
\colhead{Field}&\colhead{MCP} &\colhead{IR} &\colhead{4E}&\colhead{X-ray}&&\colhead{Cloud}&\colhead{MPCM}&\colhead{Known} \\ \cline{2-4} \cline{6-8}
\colhead{(1)} &  \colhead{(2)} & \colhead{(3)} & \colhead{(4)} & \colhead{(5)} && \colhead{(6)} & \colhead{(7)} & \colhead{(8)} 
}
\startdata 
W3          & 41~~  & 14~~  & 1~~   & 27~~~ && 39~~~        & 37~~~ & 15~~~~    \\ 
Flame Nebula    & 23~~  & 12~~  & 6~~   & 8~~~  && 16~~~    & 15~~~     & 24~~~~    \\
Rosette Nebula  & 67~~  & 60~~  & 8~~   & 5~~~  && 53~~~        & 44~~~ & 32~~~~    \\
NGC 2264    & 74~~  & 37~~  & 43~~  & 11~~~ && 72~~~        & 62~~~ & 14~~~~    \\ 
RCW 38      & 85~~  & 69~~  & 16~~  & 11~~~ && 80~~~        & 30~~~ & 23~~~~    \\
RCW 36      & 5~~   & 2~~   & 1~~   & 3~~~  && 5~~~     & 4~~~  & 0~~~~ \\
NGC 6334    & 146~~ & 112~~ & 42~~  & 16~~~ && 140~~~   & 57~~~ & 17~~~~    \\
NGC 6357    & 84~~  & 60~~  & 14~~  & 17~~~ && 74~~~        & 41~~~ & 11~~~~    \\
Trifid Nebula   & 54~~  & 40~~  & 12~~  & 8~~~  && 50~~~        & 17~~~ & 18~~~~    \\
Lagoon Nebula   & 62~~  & 29~~  & 5~~   & 30~~~ && 34~~~        & 40~~~ & 16~~~~    \\
M 17            & 180~~ & 92~~  & 32~~  & 80~~~ && 70~~~        & 77~~~ & 67~~~~    \\
Eagle Nebula    & 87~~  & 62~~  & 13~~  & 19~~~ && 142~~~   & 63~~~ & 29~~~~    \\
W 40            & 100~~ & 60~~  & 57~~  & 3~~~  && 93~~~        & 11~~~ & 15~~~~    \\
DR 21       & 101~~ & 61~~  & 45~~  & 15~~~ && 87~~~        & 69~~~ & 28~~~~    \\ 
&&&&&&& \\
All Fields      & 1,109~~   & 710~~ & 295~~ &253~~~ && 955~~~   &567~~~ &308~~~~ \\
\enddata
\vspace{3in}
\end{deluxetable}

\newpage

\begin{deluxetable}{llllcccccrrrcrrrrc}
\tabletypesize{\tiny}
\rotate
\setlength{\tabcolsep}{0.05in} 
\tablecaption{MYStIX Candidate Protostars \label{MCP.tbl}}
\tablewidth{0pt}
\tablehead{
\colhead{MCP} & \multicolumn{5}{c}{MYStIX} & \colhead{Hers} & \colhead{Sub} & \multicolumn{4}{c}{Chandra X-ray}  &&  \multicolumn{4}{c}{Spitzer Infrared} & \colhead{Notes} \\ \cline{2-6} \cline{9-12} \cline{14-17}

 &  \colhead{Name} & \colhead{R.A.} & \colhead{Dec} & \colhead{MPCM} & \colhead{MCP} & \colhead{core} & \colhead{clus} & \colhead{ME} & \colhead{Cts} & \colhead{P(src)} & \colhead{P(var)}  &&  \colhead{[3.6]} & \colhead{[4.5]} & \colhead{[5.8]} & \colhead{[8.0]} & \\ 
 
 &  & \multicolumn{2}{c}{J2000} &  & sel  &  &  & keV &  &  &  &&  \multicolumn{4}{c}{mag} & \\
 
\colhead{(1)} & \colhead{(2)} & \colhead{(3)} & \colhead{(4)} & \colhead{(5)} & \colhead{(6)} & \colhead{(7)} & \colhead{(8)} & \colhead{(9)} & \colhead{(10)} & \colhead{(11)}  &  \colhead{(12)} && \colhead{(13)} & \colhead{(14)} & \colhead{(15)} & \colhead{(16)} & \colhead{(17)}  \\
}
\startdata 

\multicolumn{18}{c}{\bf W3} \\
MCP W3 1 & G133.6785+00.9244 & 36.12805 & 61.83859  &    $-$            & IR & $-$  & \nodata   & \nodata & \nodata & \nodata & \nodata     && 10.94 & 10.12 & 9.23 & 7.86 & $\surd$ \\
MCP W3 2 & 022440.91+615824.9 & 36.17050 & 61.97361 & $\surd$   & X & $-$       & \nodata   & 4.6 & 10.6 & 0.0007 & $>$0.1      && \nodata & \nodata & \nodata & \nodata &  \\
MCP W3 3 & G133.6461+01.1799 & 36.25519 & 62.08913 & $\surd$    & IR & $\surd$  & \nodata   & \nodata & \nodata & \nodata & \nodata     && 10.89 & 9.91 & 9.05 & 8.14 & $\surd$  \\
MCP W3 4 & G133.6262+01.2418 & 36.26205 & 62.15410 & $\surd$    & IR & $\surd$  & \nodata   & \nodata & \nodata & \nodata & \nodata     && 12.84 & 11.80 & 10.93 & 10.14 & $\surd$  \\
MCP W3 5 & G133.6292+01.2354 & 36.26335 & 62.14703 & $\surd$    & IR & $\surd$  & \nodata   & \nodata & \nodata & \nodata & \nodata     && 13.78 & 12.93 & 12.21 & 11.47 & $\surd$  \\
MCP W3 6 & 022515.09+621354.4 & 36.31288 & 62.23180 & $\surd$   & X & $\surd$   & \nodata   & 4.5 & 17.9 & $<$0.0001 & $>$0.1   && \nodata & \nodata & \nodata & \nodata & \\
MCP W3 7 & 022517.08+620750.7 & 36.32121 & 62.13076 & $\surd$   & X & $\surd$   & \nodata   & 4.6 & 11.7 & $<$0.0001 & $>$0.1   && \nodata & \nodata & \nodata & \nodata & \\
MCP W3 8 & 022529.22+620619.5 & 36.37177 & 62.10543 & $\surd$   & X & $\surd$   & \nodata   & 4.6 & 27.7 & $<$0.0001 & $>$0.1   && \nodata & \nodata & \nodata & \nodata & $\surd$ \\
MCP W3 9 & G133.7492+01.0683 & 36.37638 & 61.94830 & $\surd$    & IR & $\surd$  & \nodata   & \nodata & \nodata & \nodata & \nodata     && 13.09 & 12.18 & 11.47 & 10.83 & $\surd$  \\
MCP W3 10 & G133.7487+01.0706 & 36.37704 & 61.95056 & $\surd$   & IR & $\surd$  & \nodata   & \nodata & \nodata & \nodata & \nodata     && 14.23 & 12.29 & 10.91 & 10.04 &  $\surd$ \\
\enddata
\tablecomments{
See \S\ref{present.sec} for column explanations. The presence of an object note in Table \ref{MCPnotes.tbl} is given in column 17. Table \ref{MCP.tbl} with 1,109 MYStIX Candidate Protostars is published in its entirety in the machine-readable format. A portion is shown here for guidance regarding its form and content.
}
\end{deluxetable}


\begin{deluxetable}{ll}
\tablecaption{Notes on individual MYStIX Candidate Protostars \label{MCPnotes.tbl}}
\tablewidth{0pt}
\tablehead{\colhead{MCP} & \colhead{Notes}}
\startdata
MCP W3 1  & 2MASS J02243079+6150189, K=12.5, Class 0/I \citep{Rivera-Ingraham11} \\
MCP W3 3  & 2MASS J02250127+6205209, Class 0/I  \citep{Rivera-Ingraham11} \\
MCP W3 4  & 2MASS J02250292+6209149, Class 0/I  \citep{Rivera-Ingraham11} \\
MCP W3 5  & 2MASS J02250325+6208493, Class II  \citep{Rivera-Ingraham11}  \\
MCP W3 8  & W3 D$_3$.  Variable X-ray source within H~II region D \citep{Hofner02}. \\
MCP W3 9  & [RMP2011] J022530.41+615653.88, Class II \citep{Rivera-Ingraham11} \\
MCP W3 10 & [RMP2011] J022530.58+615702.16, Class 0/I \citep{Rivera-Ingraham11} \\
\enddata
\tablecomments{Table \ref{MCPnotes.tbl} with notes for 415 individual MYStIX Candidate Protostars is published in its entirety in the machine-readable format. A portion is shown here for guidance regarding its form and content.}
\end{deluxetable}

\end{document}